# DXA-Derived Skeletal Phenotypes and Hip Fracture Risk: A Backdoor-Adjusted Causal Analysis

Zixin Shi[1], Chen Zhao[2], Meiling Zhou[3], Kevin A Maupin[4], Joyce H Keyak[5], Nancy E Lane[6], Kuan-Jui Su[7], Hui Shen[7], Hong-Wen Deng[7], Kui Zhang[8], Weihua Zhou[1,9]*

[1] Department of Applied Computing, Michigan Technological University, Houghton, MI, 49931, USA

[2] Department of Computer Science, Kennesaw State University, Marietta, GA, 30060, USA

[3] Department of Statistics, Grand Valley State University, Allendale, MI, USA

[4] Department of Information Sciences and Technologies, Grand Valley State University, Allendale, MI, USA

[5] Department of Radiological Sciences, Department of Biomedical Engineering, and Department of Mechanical and Aerospace Engineering, University of California, Irvine, CA, USA

[6] Department of Internal Medicine and Division of Rheumatology, UC Davis Health, Sacramento, CA 95817

[7] Division of Biomedical Informatics and Genomics, Tulane Center of Biomedical Informatics and Genomics, Deming Department of Medicine, Tulane University, New Orleans, LA 70112

[8] Department of Mathematical Sciences, Michigan Technological University, Houghton, MI, USA

[9] Center for Biocomputing and Digital Health, Institute of Computing and Cybersystems, and Health Research Institute, Michigan Technological University, Houghton, MI, 49931, USA

* Corresponding author:

Weihua Zhou, Ph.D.

E-Mail: whzhou@mtu.edu

Department of Applied Computing, Michigan Technological University, Houghton, MI, USA

**Abstract:**
**Purpose:** To compare dual-energy X-ray absorptiometry (DXA)-derived hip skeletal phenotypes in relation to hip fracture risk using prespecified confounder adjustment and to assess whether phenotypes ranked by their backdoor-adjusted average treatment effects (ATEs) improve risk stratification.
**Methods:** We analyzed 21,098 UK Biobank participants with linked health records, hip DXA-derived skeletal measures, and prespecified covariates. Sixteen phenotypes spanning bone mineral content (BMC), bone mineral density (BMD), and T-score across hip-related regions were evaluated. Confounder selection was guided by a prespecified directed acyclic graph (DAG). Backdoor-adjusted ATEs were estimated on the absolute risk-difference scale per standard deviation (SD) increase. Effect heterogeneity was evaluated for total femur BMD, and downstream prediction was assessed using clinical variables combined with phenotypes ranked by ATE magnitude.
**Results:** Among 21,098 participants, 115 had hip fractures. All 16 phenotypes showed negative backdoor-adjusted ATEs per SD increase. The largest ATEs were observed for total femur BMC and total femur BMD, each with a risk difference of −0.0047, corresponding to approximately 4.7 fewer hip fractures per 1,000 participants per SD higher phenotype value. Conditional effects of total femur BMD were stronger among older participants and those with lower BMI. In prediction, clinical variables plus the top 11 ATE-ranked phenotypes achieved higher AUC than FRAX with femoral neck BMD (0.842 vs. 0.709), with higher sensitivity (0.748 vs. 0.443) and similar specificity (0.793 vs. 0.777).
**Conclusion:** DXA-derived hip skeletal phenotypes differed in their backdoor-adjusted ATEs. Phenotype-level causal evaluation may help identify informative DXA measures for risk stratification.

## 1.Introduction

Hip fracture is one of the most devastating and burdensome health events in older adults [1]. In the United States, approximately 300,000 people experience hip fractures each year, with annual treatment expenditures exceeding $17 billion [2]. First-year mortality after hip fracture remains high worldwide, ranging from 12% to 36% [3], and survivors often experience substantial disability [4], loss of independence, and health complications [5]. Because of these consequences, accurate identification of individuals at increased hip fracture risk remains an important clinical and public health priority.

Dual-energy X-ray absorptiometry (DXA)-derived areal bone mineral density (aBMD) and T-score are central tools for osteoporosis assessment and fracture-risk evaluation [6, 7]. The Fracture Risk Assessment Tool (FRAX) is also widely used to estimate 10-year fracture probability based on clinical risk factors [8], with optional inclusion of femoral neck BMD. However, hip fracture risk is influenced by multiple skeletal, clinical, and functional factors. Conventional DXA-based assessment and FRAX are commonly used for clinical fracture-risk assessment, but they rely primarily on clinical risk factors and selected summary DXA measures rather than a systematic comparison of multiple region-specific hip skeletal phenotypes [9, 10]. Because DXA contains several region-specific skeletal measurements beyond the conventional summary measures commonly used in clinical practice, a systematic evaluation of these phenotypes may provide additional insight into hip fracture risk.

Large observational cohorts such as the UK Biobank provide an opportunity to examine DXA-derived skeletal phenotypes together with demographic, lifestyle, clinical, and functional risk factors [11]. Previous studies have largely focused on fracture-risk prediction, associational analyses [12], or genetically informed causal inference [13]. Machine-learning and imaging-based studies have improved fracture-risk discrimination [14, 15], and Mendelian randomization studies have provided important evidence regarding genetically predicted skeletal traits and fracture risk [13, 16]. However, these approaches do not directly compare multiple DXA-derived skeletal phenotypes within an observational cohort under an explicit confounder-adjustment framework [17, 18]. This distinction is important because predictors that improve risk classification are not necessarily the same variables that show larger backdoor-adjusted ATEs [19]. A prespecified causal framework can therefore help clarify which DXA-derived phenotypes show stronger adjusted relationships with hip fracture risk under explicit assumptions about confounding and identification [20, 21].

In this study, we used UK Biobank cohort to evaluate 16 DXA-derived hip skeletal phenotypes across anatomical regions and measurement families in relation to hip fracture risk. We applied a prespecified directed acyclic graph (DAG)-guided backdoor-adjustment framework to estimate backdoor-adjusted average treatment effects (ATEs) on the absolute risk-difference scale. We further examined conditional average treatment effects (CATEs) for the leading phenotype and assessed whether phenotypes ranked by the magnitude of their backdoor-adjusted ATEs improved downstream risk stratification when combined with clinical variables. The aim was to identify DXA-derived skeletal phenotypes with larger backdoor-adjusted ATEs and potential clinical utility for more parsimonious risk stratification.

## 2.Materials and methods

### 2.1 Study population and descriptive analysis

The observational cohort study used data from the UK Biobank under Application No. 61915. Participants were eligible for inclusion if they had available linked health record data, DXA-derived skeletal measurements, and prespecified covariates required for the backdoor causal analysis. Hip fracture was defined using linked health records based on ICD-10 codes S72.0, S72.1, and S72.2 [22]. Participants with missing values in any prespecified DXA-derived skeletal phenotype, outcome, or adjustment covariate were excluded. The final complete-case analytic cohort was used for the primary causal analysis and downstream predictive evaluation.

Baseline characteristics were summarized according to hip fracture status. Continuous variables were reported as mean ± standard deviation (SD) and compared between groups using the Mann-Whitney U test [23]. Categorical variables were summarized as counts and percentages and compared using the chi-square test [24] or Fisher's exact test [25], as appropriate. Standardized mean differences (SMDs) [26] were also calculated to quantify the magnitude of between-group differences. These baseline comparisons were used to describe the analytic cohort and were not used for covariate selection. Adjustment variables were prespecified based on clinical knowledge, prior literature, and the DAG-guided causal framework described below.

## 2.2 DXA-derived skeletal phenotypes and DAG-guided confounder selection

In this study, hip fracture was defined as the outcome, and DXA-derived skeletal phenotypes were treated as the exposures of interest. We prespecified a subject-matter-informed DAG [27] based on prior literature and clinical knowledge to represent the hypothesized causal relationships among DXA-derived skeletal phenotypes, prespecified confounders, and the hip fracture. In the conceptual DAG, each DXA-derived skeletal phenotype $X_j (j = 1, \ldots, 16)$was evaluated separately as the exposure in a phenotype-specific analysis, hip fracture was treated as the outcome $Y$, and the prespecified confounder set $C$was selected to block major backdoor paths between $X_j$and $Y$. The graphical DAG is provided in Supplementary Figure S1.

The exposure set consisted of 16 prespecified DXA-derived skeletal phenotypes spanning major hip-related anatomical regions, including the femoral neck, total femur, trochanteric region, Ward's region, femoral shaft, and pelvis, and measurement families, including bone mineral content (BMC), bone mineral density (BMD), and T-score. These phenotypes were selected a priori to provide systematic coverage of available hip skeletal measures in the UK Biobank and to allow comparative evaluation across anatomical regions and measurement families.

Candidate confounders were selected a priori based on the prespecified DAG, prior literature, and clinical knowledge. Variables were included if they were considered plausible common causes, or clinical proxies of common causes, of both DXA-derived skeletal phenotypes and hip fracture. Age and sex were included because they are strongly related to both skeletal phenotype variation and fracture risk [28-30]. Height and weight were included as body-size factors related to DXA-derived skeletal measurements, skeletal loading, and fracture risk [31-33]. Household income was included as a socioeconomic indicator associated with bone health and fracture risk [34]. Smoking status, alcohol use, and major dietary change were included as lifestyle and nutritional factors related to skeletal health and fracture susceptibility [35-38]. Rheumatoid arthritis, falls in the last year, and prior fractures were included as clinical risk factors associated with bone loss, falls, or future fracture risk [19, 39-41]. Usually walking pace and hand grip strength were included as functional indicators related to frailty, mobility, muscle strength, and fracture risk [42, 43]. This approach was intended to reduce confounding rather than to select variables based on statistical significance. The overall DAG-guided backdoor-adjustment workflow is summarized in Figure 1. Briefly, each DXA-derived skeletal phenotype was evaluated as the exposure in a separate phenotype-specific model, hip fracture was treated as the outcome, and the prespecified confounder set was used to block major backdoor paths of the form X ← C → Y. The DAG-guided analytic specification of the outcome, exposures, and prespecified confounders is summarized in Table 1.

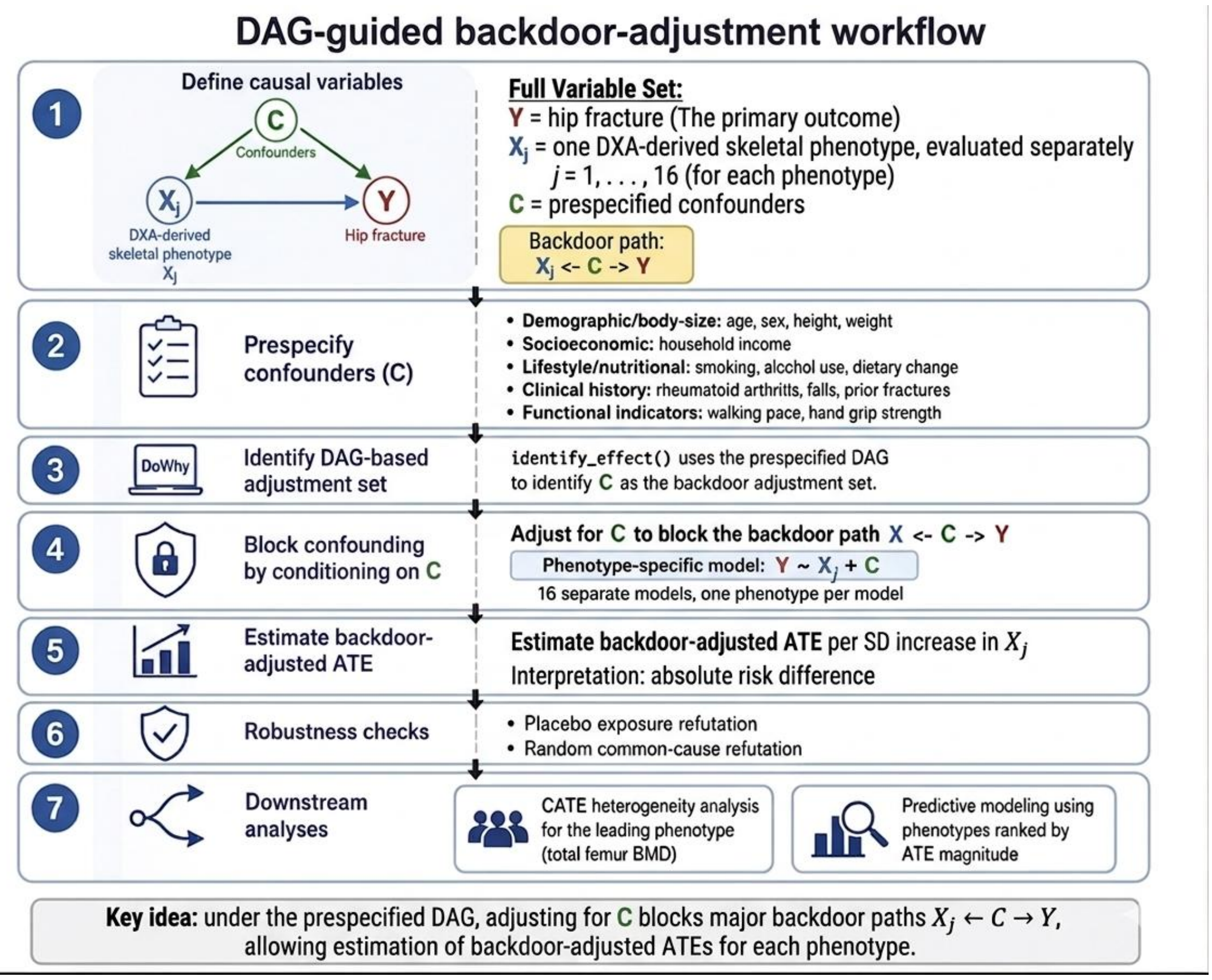

Figure 1. Overview of the DAG-guided backdoor-adjustment workflow.

Table 1. DAG-guided analytic specification of the outcome, exposures, and prespecified confounders. In the conceptual DAG, each DXA-derived skeletal phenotype was treated as an exposure $X$, hip fracture was treated as the outcome $Y$, and the listed variables were treated as prespecified confounders $C$. Confounders were selected to block major backdoor paths of the form $X \leftarrow C \rightarrow Y$. BMC, bone mineral content; BMD, bone mineral density; DAG, directed acyclic graph; DXA, dual-energy X-ray absorptiometry.

| DAG component | Domain / family | Variable(s) | Role or rationale in the DAG-guided analysis |
|---|---|---|---|
| Outcome, Y | Hip fracture | Hip fracture defined from linked health records using ICD-10 codes S72.0, S72.1, and S72.2 | Outcome of interest |
| Exposure, X | BMC | Total femur BMC; femoral neck BMC; femoral shaft BMC; trochanteric region BMC; Ward's region BMC; pelvis BMC | Each phenotype was evaluated separately as a continuous exposure |
| | BMD | Total femur BMD; femoral neck BMD; femoral shaft BMD; trochanteric region BMD; Ward's region BMD; pelvis BMD | |
| | T-score | Total femur T-score; femoral neck T-score; | |

| | | trochanteric region T-score; Ward's region T-score | |
|---|---|---|---|
| Confounder, C | Demographic/body-size factors | Age; sex; height; weight | Prespecified adjustment variables selected to block major backdoor paths between X and Y |
| | Socioeconomic status | Household income | Prespecified adjustment variable selected to block major backdoor paths between X and Y |
| | Lifestyle/nutritional factors | Smoking status; alcohol use; major dietary change in the last 5 years | Prespecified adjustment variables selected to block major backdoor paths between X and Y |
| | Clinical history | Rheumatoid arthritis; falls in the last year; prior fractures | Prespecified adjustment variables selected to block major backdoor paths between X and Y |
| | Functional indicators | Usually walking pace; hand grip strength | Prespecified adjustment variables selected to block major backdoor paths between X and Y |

## 2.3 Backdoor-adjusted average treatment effect estimation

As shown in Figure 1, the primary causal analysis used phenotype-specific backdoor-adjusted models rather than a joint model containing all 16 skeletal phenotypes simultaneously. For each of the 16 prespecified DXA-derived skeletal phenotypes, we estimated the backdoor-adjusted ATE on hip fracture risk using the complete-case analytic cohort. Under the assumptions represented by the prespecified DAG, the adjustment set was used to block major backdoor paths between each skeletal phenotype and hip fracture. If the prespecified confounder set $C$blocks these backdoor paths between a DXA-derived skeletal phenotype $X$and hip fracture $Y$, the interventional mean outcome can be identified using the backdoor adjustment formula [44, 45], as shown in Eq. 1.

$$E[Y|do(X=x)] = \sum_{c} E[Y|X=x, C=c]P(C=c) \qquad (1)$$

where $Y$denotes hip fracture status, $X$denotes a given DXA-derived skeletal phenotype, $x$denotes a specified value of that phenotype $X$, $C$denotes the prespecified confounder set, and $c$denotes a set of specific values of all covariates. This expression represents the expected hip fracture risk under a hypothetical intervention setting the skeletal phenotype to value $x$, averaged over the distribution of the prespecified confounders.

The causal interpretation of the backdoor-adjusted ATEs relies on standard assumptions for observational causal inference, including consistency, positivity, no interference, correct specification of the prespecified DAG and adjustment set, and absence of residual or unmeasured confounding after adjustment for the selected covariates. Under these assumptions, conditioning on the prespecified confounder set blocks major backdoor paths between each DXA-derived skeletal phenotype and hip fracture.

Categorical covariates were represented using indicator variables before modeling. For the primary analysis, we used a linear probability model with an identity link for regression-based backdoor adjustment, fitting one DXA-derived skeletal phenotype and the prespecified confounders in each phenotype-specific model [44, 45]. Because hip fracture was a binary outcome, backdoor-adjusted

ATEs were interpreted on the absolute risk-difference scale. In this setting, the ATE represents the average change in hip fracture probability associated with a higher value of a given DXA-derived skeletal phenotype after adjustment for the prespecified confounder set.

To facilitate comparison across skeletal phenotypes measured on different scales, the primary results were reported by per standard deviation (SD) increase in each phenotype. Original-unit ATEs were also obtained but were considered secondary because the clinical meaning of a one-unit increase differs across BMC, BMD, and T-score measures. Under the adjusted linear regression model, the per-SD ATE was obtained by multiplying the fitted exposure coefficient by the sample SD of the corresponding skeletal phenotype. Ninety-five percent confidence intervals were derived from adjusted linear regression models fitted using the same exposure and confounder specification as the primary analysis. Robustness was evaluated using placebo exposure and random common-cause refutation analyses. In the placebo exposure refutation, the original skeletal phenotype was replaced with a randomly generated exposure. In the random common-cause refutation, a randomly generated covariate was added to assess whether the ATEs were materially changed by a non-informative covariate.

## 2.4 Heterogeneity of conditional average treatment effects

We further examined whether CATEs for DXA-derived skeletal phenotypes differed across participants with different clinical and demographic profiles. Each skeletal phenotype was modeled separately as a continuous exposure, with hip fracture as the outcome. The same prespecified covariates used in the primary backdoor-adjusted analysis were included to account for baseline differences and to characterize potential effect heterogeneity.

To quantify heterogeneity, we estimated CATEs. In this study, CATEs represent the conditional effect of a DXA-derived skeletal phenotype on hip fracture risk conditional on an individual participant's observed covariate profile. Because hip fracture was a binary outcome, CATEs were interpreted as conditional risk differences in the estimated probability of hip fracture. To facilitate comparison across skeletal phenotypes measured on different scales, CATEs were estimated per SD increase in each phenotype.

For this analysis, skeletal phenotype values were standardized by their sample SD. The analytic cohort was divided into training and test sets using 80:20 stratified sampling by hip fracture status. Causal forest models were used to estimate individual-level and subgroup-level variation in the adjusted exposure–outcome relationship. These models allow treatment effects to vary across covariate profiles rather than assuming a single constant effect for all participants [46, 47].

The resulting CATEs were summarized using the distribution of predicted individual effects, the relative importance of covariates contributing to heterogeneity, and subgroup summaries by age, sex, and body mass index. Body mass index was recalculated from height and weight for subgroup analyses. In the main text, we focused on femur total bone mineral density because it showed one of the largest backdoor-adjusted ATEs in the primary analysis. Results for the remaining skeletal phenotypes and detailed model parameters are provided in the Supplementary Methods and Supplementary Figures.

## 2.5 Predictive modeling and model interpretation

To evaluate whether DXA-derived skeletal phenotypes ranked by the magnitude of their backdoor-adjusted ATEs provided additional value for hip fracture risk stratification, we performed a supplementary predictive analysis. Logistic regression was used as the primary prediction model because it provided good discrimination while maintaining clinical interpretability. To address the imbalance between hip fracture cases and non-cases, class-weighted learning was applied. Class weights were assigned inversely proportional to class frequencies in the training data; during cross-validation, the weights were recalculated within each training fold so that the held-out fold was not used to define class weights.

We first examined whether predictive performance improved as additional ATE-ranked skeletal phenotypes were added. In the top-K analysis, K denoted the number of skeletal phenotypes included according to their ATE-based ranking. Skeletal phenotypes were ranked according to the magnitude of their backdoor-adjusted ATEs, and the top-K phenotypes were added sequentially from K = 1 to K = 16.

Predictive performance was assessed using 5-fold stratified cross-validation, with the area under the receiver operating characteristic curve (AUC) used as the primary discrimination metric.

We then compared predefined feature-set configurations to evaluate the incremental value of DXA-derived skeletal phenotypes beyond clinical variables alone. The compared feature sets included clinical variables only, selected top-K ATE-ranked skeletal phenotypes only, all skeletal phenotypes, all available clinical and skeletal features, and clinical variables combined with the selected top-K ATE-ranked skeletal phenotypes. The selected value of K was determined from the top-K analysis and subsequent feature-set comparison. Categorical variables were encoded as indicator variables before model fitting. The predictive analysis was conducted using the same complete-case analytic cohort as the primary analysis. Model performance was summarized using AUC, accuracy, sensitivity, and specificity.

Finally, model interpretation was performed to characterize the relative contribution of clinical and skeletal features to the final prediction model. Feature contribution analyses were used to assess whether the final model was driven primarily by DXA-derived skeletal phenotypes, clinical variables, or their combination. This predictive analysis was considered complementary to the primary backdoor-adjusted ATE analysis and was intended to evaluate the potential value of ATE-ranked skeletal phenotypes for parsimonious hip fracture risk stratification.

## 3.Results

### 3.1 Study population and baseline characteristics

The final complete-case analytic cohort included 21,098 UK Biobank participants, including 115 participants with hip fractures and 20,983 participants without hip fractures. Baseline characteristics according to hip fracture status are summarized in Table 2. Participants with hip fracture were older than those without hip fracture, with a mean age of 69.43 years versus 63.94 years. The hip fracture group also had a higher proportion of female participants, with 71 of 115 participants being female (61.7%), compared with 10,665 of 20,983 participants in the non-fracture group (50.83%). Mean body weight was 71.68 kg in the hip fracture group and 75.96 kg in the non-fracture group. Differences were also observed in household income distribution, alcohol use, rheumatoid arthritis history, usual walking pace, and hand grip strength. The largest between-group differences, as assessed by standardized mean differences, were observed for rheumatoid arthritis, age, household income, alcohol use, and hand grip strength. In contrast, smoking status, dietary change, falls in the previous year, and prior fracture history showed smaller between-group differences. These baseline comparisons describe the analytic cohort; covariates for adjustment were prespecified based on the DAG-guided causal framework rather than selected from baseline comparisons.

Table 2. Baseline characteristics of the complete-case analytic cohort by hip fracture status. Values are presented as mean ± SD for continuous variables and count (n) (percentage - %) for categorical variables. One alcohol unit corresponds to 10 mL or 8 g of pure alcohol. SMD, standardized mean difference; SD, standard deviation.

| Characteristic | | No hip fracture (n = 20,983) | Hip fracture (n = 115) | SMD |
|---|---|---|---|---|
| Age, years, mean ± SD | | 63.94 ± 7.73 | 69.43 ± 6.57 | 0.712 |
| Female, n (%) | | 10665 (50.83%) | 71 (61.70%) | 0.22 |
| Male, n (%) | | 10318 (49.17%) | 44 (38.30%) | |
| Height, cm, mean ± SD | | 170.66 ± 9.33 | 169.39 ± 9.74 | 0.136 |
| Weight, kg, mean ± SD | | 75.96 ± 15.01 | 71.68 ± 13.88 | 0.285 |
| Household income, n (%) | Less than 18k | 2148 (10.24%) | 39 (33.91%) | 0.624 |
| | 18k to 31k | 5479 (26.11%) | 23 (20.00%) | |
| | 31k to 52k | 6549 (31.21%) | 27 (23.48%) | |
| | 52k to 100k | 5075 (24.19%) | 22 (19.13%) | |
| | Greater than 100k | 1732 (8.25%) | 4 (3.48%) | |
| Smoking status, n (%) | Never | 12946 (61.70%) | 67 (58.26%) | 0.07 |
| | Previous | 7351 (35.03%) | 44 (38.26%) | |

| | Current | 686 (3.27%) | 4 (3.48%) | |
|---|---|---|---|---|
| Alcohol use, units/day, n (%) | 1 or 2 | 9994 (47.63%) | 70 (60.87%) | 0.326 |
| | 3 or 4 | 6187 (29.49%) | 27 (23.48%) | |
| | 5 or 6 | 2742 (13.07%) | 13 (11.30%) | |
| | 7, 8 or 9 | 1472 (7.02%) | 4 (3.48%) | |
| | 10 or more | 588 (2.80%) | 1 (0.87%) | |
| Dietary change, n (%) | No | 13569 (64.67%) | 71 (61.74%) | 0.062 |
| | Yes | 7414 (35.33%) | 44 (38.26%) | |
| Rheumatoid arthritis, n (%) | Yes | 1056 (5.03%) | 31 (26.96%) | 0.73 |
| | No | 19927 (94.97%) | 84 (73.04%) | |
| Falls, n (%) | No falls | 17401 (82.93%) | 89 (77.39%) | 0.149 |
| | = one fall | 2683 (12.79%) | 18 (15.65%) | |
| | > one fall | 899 (4.28%) | 8 (6.96%) | |
| Prior fractures, n (%) | No | 19173 (91.37%) | 105 (91.30%) | 0.002 |
| | Yes | 1810 (8.63%) | 10 (8.70%) | |
| Usually walking pace, n (%) | Slow pace | 794 (3.78%) | 12 (10.43%) | 0.277 |
| | Steady average pace | 10054 (47.91%) | 56 (48.70%) | |
| | Brisk pace | 10135 (48.30%) | 47 (40.87%) | |
| Hand grip strength, kg, mean ± SD | | 30.36 ± 10.21 | 27.23 ± 10.14 | 0.307 |

### 3.2 Backdoor-adjusted ATEs of DXA-derived skeletal phenotypes on hip fracture risk

Using the prespecified DAG-guided backdoor-adjustment framework, all 16 DXA-derived skeletal phenotypes showed negative backdoor-adjusted ATEs per SD increase, indicating lower expected hip fracture probability after adjustment for the prespecified confounders (Figure 2; Supplementary Table S1). The largest per-SD ATEs were observed for total femur BMC and total femur BMD, each with a risk difference of −0.0047. Given the baseline hip fracture risk of 115/21,098, or approximately 5.45 cases per 1,000 participants, this ATE corresponds to approximately 4.7 fewer hip fractures per 1,000 participants per SD higher phenotype value in this analytic cohort. Ward's region BMC showed the smallest ATE (−0.0019). The 95% confidence intervals for all 16 skeletal phenotypes excluded zero.

The inverse ATEs varied by feature family and skeletal region (Figure 3). Total femur measures showed the largest ATEs across feature families, whereas Ward's region BMC and pelvis-based measures showed comparatively smaller ATEs. BMD-based measures showed relatively strong and consistent inverse ATEs across regions. These patterns suggest that DXA-derived skeletal phenotypes differed in their backdoor-adjusted ATEs rather than contributing uniformly across regions and measurement families. Refutation analyses supported the stability of the primary findings (Supplementary Figure S2): placebo exposure ATEs were centered near zero, and random common-cause ATEs remained close to the original ATEs without reversal of direction.

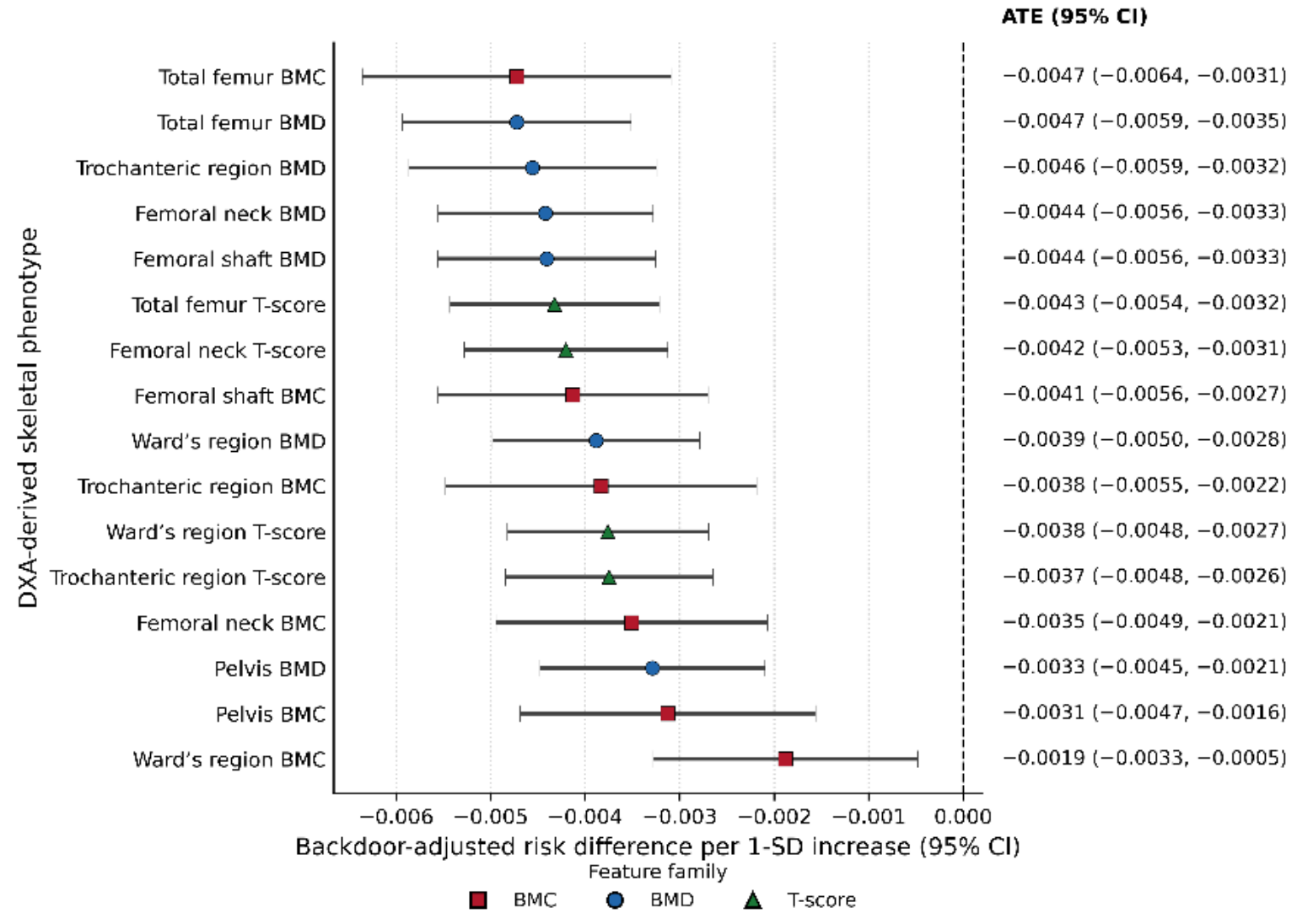

Figure 2. Backdoor-adjusted ATEs of DXA-derived skeletal phenotypes on hip fracture risk per SD increase.

Darker blue indicates a stronger inverse adjusted estimate. Blank cells indicate unavailable DXA-derived measures.

Figure 3. Heatmap of backdoor-adjusted ATEs by skeletal region and feature family.

### 3.3 Heterogeneity in CATEs for total femur BMD

As a complementary analysis, we examined whether CATEs for total femur BMD varied across participants with different demographic and clinical profiles. Although heterogeneity analyses were conducted for all 16 DXA-derived skeletal phenotypes, total femur BMD is presented in the main text because it was among the phenotypes with the largest backdoor-adjusted ATEs in the primary analysis. Results for the remaining phenotypes are provided in Supplementary Figures S3–S11. As shown in Figure 4A–I, predicted CATEs for total femur BMD were not uniform across individuals, indicating that the conditional effect of higher total femur BMD differed across participant profiles.

Feature-importance analysis identified household income as the strongest contributor to CATE heterogeneity, followed by age, weight, rheumatoid arthritis, and hand grip strength, whereas smoking, walking pace, and dietary change contributed less prominently (Figure 4A). Age-related heterogeneity was observed in both the age-sex CATE curves and the age-group summaries. The predicted CATEs were generally negative across most of the age range for both sex groups, with more negative CATEs at older ages and comparatively small differences between women and men (Figure 4B and 4E). In the subgroup summaries, mean CATEs became progressively more negative across older age groups, indicating more negative CATEs for total femur BMD among older participants (Figure 4D). CATEs also varied by BMI group, with lower-BMI groups showing more negative mean CATEs than higher-BMI groups, suggesting a stronger CATEs among participants with lower BMI (Figure 4F). The decile-based summary further supported heterogeneity, with the most negative mean CATE observed in the lowest predicted-CATE decile and progressively less negative CATEs across higher deciles (Figure 4C).

Individual-level analyses further illustrated variation in the predicted CATEs for total femur BMD (Figure 4G–I). The distribution of individual predicted CATEs was centered below zero and showed a left tail toward more negative values, indicating that some participants had larger predicted risk reductions than others (Figure 4G). The ranking curve showed a continuous gradient from more negative to near-null predicted CATEs, rather than a uniform effect across all participants (Figure 3H). Similarly, the top-benefit population analysis showed that the average predicted CATE was most negative among participants with the largest predicted benefit and became less negative as broader portions of the population were included (Figure 4I). Overall, these findings suggest that CATEs for total femur BMD varied across participant profiles, with more negative CATEs observed particularly among older participants, lower BMI participants, and selected high-effect subgroups.

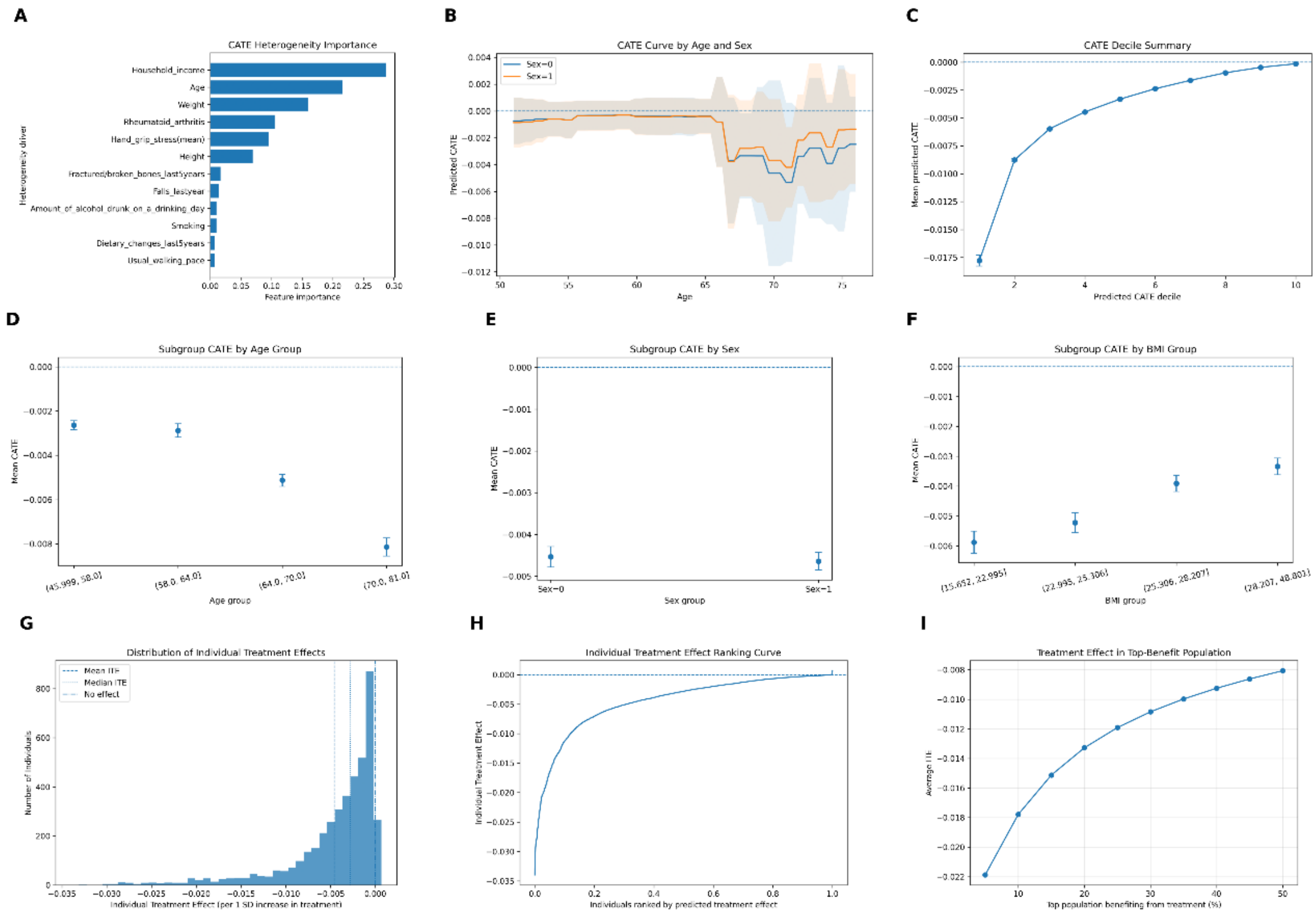

Figure 4. Heterogeneity in CATEs for total femur BMD on hip fracture risk. (A) Variable importance for CATE heterogeneity. (B) Predicted CATE curves by age and sex. (C) Decile-based summary of mean predicted CATE. (D) Mean CATE by age group. (E) Mean CATE by sex group. (F) Mean CATE by BMI group. (G) Distribution of individual predicted CATEs. (H) Ranking curve of individual predicted CATEs. (I) Average predicted CATE in the top-benefit population. CATE, conditional average treatment effect; BMD, bone mineral density.

### 3.4 Downstream predictive evaluation

As a supplementary analysis, we evaluated whether DXA-derived skeletal phenotypes ranked by the magnitude of their backdoor-adjusted ATEs improved hip fracture risk stratification when combined with clinical variables. Logistic regression was used for the main predictive evaluation because it provided good discrimination while remaining clinically interpretable. Full model-screening results for the evaluated classifiers are provided in Supplementary Table S2.

We first examined whether predictive performance improved as additional ATE-ranked skeletal phenotypes were added. Skeletal phenotypes were ranked according to the magnitude of their backdoor-adjusted ATEs and added sequentially from 1 to 16. Predictive performance increased most noticeably after the first few ATE-ranked skeletal phenotypes were added. Here, K denotes the number of top-ranked skeletal phenotypes included in the model. In the full-cohort ATE ranking, K = 1 corresponded to total femur BMC, and K = 2 added total femur BMD. Cross-validated AUC increased from approximately 0.665 at K = 1 to 0.787 at K = 2. Performance then increased more gradually and reached a broad plateau around 0.81, with near-maximal performance observed from approximately K = 11 onward. Based on this plateau pattern and the subsequent feature-set comparison, the top 11 ATE-ranked skeletal phenotypes were selected for downstream predictive evaluation. The full top-K analysis is shown in Supplementary Figure S12.

We then compared FRAX with five logistic-regression-based feature-set configurations: clinical variables only, top 11 ATE-ranked skeletal phenotypes only, all 16 skeletal phenotypes, all available clinical and skeletal features, and clinical variables combined with the top 11 ATE-ranked skeletal phenotypes (Table 3). The clinical-plus-top-11 model achieved the highest test AUC (0.842), followed closely by the all-features model (0.837). The top-11-only and all-skeletal-phenotype models also showed higher AUCs than the clinical-only and FRAX models. Compared with FRAX, the clinical-plus-top-11 model showed higher sensitivity (0.748 vs. 0.443) with similar specificity (0.793 vs. 0.777).

These findings suggest that a small set of ATE-ranked DXA-derived skeletal phenotypes may improve risk stratification when combined with clinical variables.

Model interpretation using SHAP is provided in Supplementary Figure S13. In the final logistic regression model, the largest feature contributions were mainly from DXA-derived skeletal phenotypes, including total femur BMC, trochanteric region BMC, total femur BMD, femoral neck BMD, and femoral shaft BMD. Clinical variables contributed more modestly in comparison. These SHAP results indicate that the final predictive model was driven primarily by DXA-derived skeletal phenotypes, with clinical variables providing additional but smaller contributions.

Table 3. Test-set predictive performance of FRAX and logistic-regression-based feature-set configurations. LR, logistic regression; AUC, area under the receiver operating characteristic curve; FRAX, Fracture Risk Assessment Tool.

| Configuration | Model | Test | | | |
|---|---|---|---|---|---|
| | | AUC | Accuracy | Sensitivity | Specificity |
| FRAX | FRAX | 0.709 | 0.775 | 0.443 | 0.777 |
| Clinical-only | LR | 0.754 | 0.733 | 0.626 | 0.734 |
| Top11-Causal-only | LR | 0.813 | 0.721 | 0.783 | 0.721 |
| All-causal | LR | 0.801 | 0.728 | 0.757 | 0.727 |
| All-features | LR | 0.837 | 0.797 | 0.713 | 0.797 |
| Clinical-Top11-Causal | LR | 0.842 | 0.793 | 0.748 | 0.793 |

## 4.Discussion

### Major findings

In this study, we used a prespecified DAG-guided backdoor-adjustment framework to evaluate DXA-derived hip skeletal phenotypes in relation to hip fracture risk in the UK Biobank. All 16 prespecified skeletal phenotypes showed negative backdoor-adjusted ATEs per SD increase, suggesting lower estimated hip fracture probability after adjustment for demographic, lifestyle, clinical, and functional confounders. The strongest inverse ATEs were observed for total femur BMC and total femur BMD. Heterogeneity analysis further suggested that CATEs for total femur BMD were not uniform across participants, with more negative CATEs observed among older individuals, participants with lower BMI, and selected high-effect subgroups. In supplementary predictive evaluation, combining clinical variables with the top 11 ATE-ranked skeletal phenotypes achieved the highest discrimination in this analytic cohort and outperformed FRAX. Together, these findings suggest that DXA-derived skeletal phenotypes may differ in their relevance to hip fracture risk and that phenotype-level evaluation may provide additional information for risk stratification beyond conventional summary measures alone.

### Comparison with prior fracture-risk and causal studies

These findings should also be interpreted in the context of prior hip fracture research, which has largely emphasized fracture-risk prediction, imaging-based model development, or genetically informed causal inference. Prior prediction studies have shown that imaging-derived skeletal features can improve fracture-risk assessment. Hsieh et al. used routine radiographs to develop a deep-learning pipeline for fracture detection, BMD estimation, and fracture-risk assessment [48]. In a separate study of older women, Jaiswal et al. found that adding HR-pQCT-derived bone parameters to hip fracture prediction models modestly improved discrimination, with the AUC increasing from 0.71 to 0.75 [49]. These studies support the value of skeletal imaging features for fracture-risk prediction, but they were primarily designed to improve classification performance rather than to compare backdoor-adjusted ATEs across specific DXA-derived skeletal phenotypes. Genetically informed causal studies have also provided important evidence regarding skeletal traits and hip fracture risk. Nethander et al. used GWAS meta-analysis and Mendelian randomization to identify causal effects of lower femoral neck BMD, Alzheimer's disease, and smoking on hip fracture risk [16]. Tobias et al. further showed that DXA-derived femoral neck width was causally related to hip fracture independently of femoral neck BMD using a genetic approach [50]. These studies provided important causal evidence, but they rely on genetically informed designs and do not directly compare multiple DXA-derived skeletal phenotypes in observational cohort data under explicit confounder adjustment.

Our study extends this literature by systematically evaluating 16 DXA-derived skeletal phenotypes within a prespecified DAG-guided backdoor-adjustment framework and by examining whether phenotypes ranked by their backdoor-adjusted ATEs also contribute to downstream risk stratification.

**Clinical interpretation of phenotype-level DXA effects**

The largest backdoor-adjusted ATEs were observed for total femur and BMD-based skeletal phenotypes, particularly total femur BMC and total femur BMD, both with per-SD ATEs around −0.0047 on the risk-difference scale, whereas Ward's region and pelvis-based measures showed comparatively smaller ATEs. This pattern suggests that DXA-derived skeletal phenotypes may not contribute uniformly to hip fracture risk. Importantly, these findings should be interpreted as complementary to, rather than a replacement for, conventional DXA interpretation. In current practice, aBMD and T-score remain cornerstone tools for osteoporosis assessment and fracture-risk evaluation [6, 7]. Our results suggest that additional phenotype-level evaluation across hip regions and measurement families may help refine the interpretation of DXA-derived skeletal information for risk stratification.

The findings also provide context for future research on T-score-based fracture-risk assessment. T-score remains a widely used and clinically important metric because it provides a simple summary measure for osteoporosis classification. In our analysis, T-score-based phenotypes showed negative backdoor-adjusted ATEs, but they were not consistently the phenotypes with the largest inverse ATEs. Total femur BMC and total femur BMD showed larger inverse ATEs than the evaluated T-score measures. This pattern suggests that region- and measurement-family-level DXA phenotypes may contain information not fully summarized by T-score alone. Future studies should evaluate whether selected regional DXA-derived phenotypes can improve risk stratification while remaining simple enough for clinical interpretation.

**Heterogeneity and individualized risk stratification**

The heterogeneity analysis suggested that CATEs for total femur BMD varied across participant profiles. More negative CATEs were observed among older individuals, participants with lower BMI, and selected high-effect subgroups. In contrast, differences between women and men were less pronounced after conditioning on total femur BMD and the prespecified covariates. Household income, age, weight, rheumatoid arthritis, and hand grip strength contributed prominently to CATE heterogeneity, suggesting that the relevance of total BMD may be shaped not only by skeletal status but also by broader clinical, functional, and socioeconomic context.

This finding is clinically relevant because hip fracture risk is inherently heterogeneous. Population-average ATEs are useful for identifying overall patterns, but they may obscure differences across clinically meaningful subgroups [47, 51]. The observation that total femur BMD showed more negative CATEs in older and lower-BMI participants suggests that the same skeletal phenotype may have different risk implications depending on the patient's profile [10, 52]. These results should not be interpreted as establishing subgroup-specific treatment rules, but they support the potential value of considering effect heterogeneity when using DXA-derived skeletal information for fracture-risk stratification.

**Predictive value of ATE-ranked skeletal phenotypes**

The supplementary predictive analysis suggested that DXA-derived skeletal phenotypes ranked by the magnitude of their backdoor-adjusted ATEs may have practical value for parsimonious risk stratification. Logistic regression using clinical variables combined with the top 11 ATE-ranked skeletal phenotypes achieved the highest test AUC and performed slightly better than the full-feature model (0.842 vs. 0.837). This finding suggests that ranking skeletal phenotypes by their backdoor-adjusted ATEs may help reduce feature burden while retaining predictive performance. The final clinical-plus-top-11 model also outperformed FRAX in this analytic cohort, with higher sensitivity and comparable specificity. However, this comparison should be interpreted cautiously. FRAX is an established clinical tool designed for broad fracture-risk assessment, whereas our model was developed and evaluated within a specific UK Biobank complete-case sample. Therefore, the predictive results should be viewed as complementary evidence supporting the potential utility of ATE-ranked DXA-derived phenotypes, rather than as evidence that the proposed model is ready to replace existing clinical risk tools [8]. External validation and calibration assessment in independent cohorts will be needed before clinical use. Model interpretation [53] further showed that the final logistic regression model was driven primarily by a small set of femoral skeletal phenotypes, including total femur BMC, trochanteric region BMC, total femur BMD, femoral neck BMD, and femoral shaft BMD. Clinical variables contributed more modestly in comparison but remained part of the best-performing model. These findings reinforce the idea that skeletal phenotypes and clinical risk factors provide complementary information for hip fracture risk stratification.

**Exploratory sex-stratified analyses**

To further examine potential sex-related differences, we performed sex-stratified backdoor-adjusted ATE analyses and presented the female- and male-specific forest plots in Supplementary Figure S12. The inverse ATE pattern between DXA-derived skeletal phenotypes and hip fracture risk was generally preserved in both women and men, suggesting that the main pattern observed in the full cohort was not driven by one sex alone. In the full cohort, total femur BMC and total femur BMD showed the largest per-SD ATEs. In women, femoral neck BMD, femoral neck T-score, total BMD, and total T-score showed relatively larger inverse ATEs, whereas in men, shaft BMD, total BMD, and total T-score ranked among the most prominent phenotypes. Although ATE directions were broadly consistent across sex strata, absolute ATE magnitudes appeared slightly larger in women than in men. Given the reduced number of hip fracture events after stratification, these findings should be interpreted as exploratory.

In parallel with the sex-stratified causal findings, we further evaluated whether predictive performance differed between women and men across logistic-regression-based feature-set configurations, as shown in Supplementary Table S3. Sex-specific evaluation has been emphasized in previous hip fracture prediction studies, and prior work has shown that prediction performance may differ between women and men depending on the model, predictors, and validation cohort [29, 54]. In the full cohort, the best-performing model was the clinical-plus-top-11 ATE-ranked model, with an AUC of 0.842. In women, the clinical-plus-top-11 ATE-ranked model achieved an AUC of 0.863, whereas in men, the best-performing configuration was the clinical-plus-top-8 ATE-ranked model, with an AUC of 0.770. Because sex stratification reduced the number of hip fracture events, especially among men, these findings should be interpreted as descriptive and exploratory rather than as evidence that sex-specific modeling improves prediction. Further validation in larger sex-specific cohorts is needed.

**Strengths, limitations, and future directions**

This study has several strengths. First, we used a prespecified DAG-guided backdoor-adjustment framework to evaluate multiple DXA-derived skeletal phenotypes within a single observational cohort. Second, we compared phenotypes across anatomical regions and measurement families, allowing a more detailed assessment of DXA-derived skeletal information than a single summary measure alone. Third, we integrated backdoor-adjusted ATE analysis, heterogeneity analysis, supplementary predictive evaluation, comparison with FRAX, and model interpretation within one analytic framework.

Several limitations should also be acknowledged. First, because this was an observational study, the causal interpretation of the backdoor-adjusted ATEs depends on the assumptions represented by the prespecified DAG, adequate covariate overlap, correct model specification, and absence of important residual or unmeasured confounding. Second, complete-case analysis may have introduced selection bias, especially if participants with complete DXA and covariate data differed systematically from those excluded because of missing data. Third, the number of hip fracture cases was small relative to the number of non-cases, which may affect the stability of backdoor-adjusted ATEs and CATEs, particularly in subgroup and heterogeneity analyses. Fourth, the predictive findings, including comparison with FRAX, were derived within the present analytic cohort and require external validation.

Future work should evaluate the transportability of these findings in independent cohorts, assess calibration and clinical utility, and examine whether ATE-ranked skeletal phenotypes improve risk stratification across different populations. Additional studies may also extend this framework to richer imaging-derived or image-mediated skeletal representations, which could provide a more detailed understanding of hip fracture risk mechanisms.

## 5. Conclusion

This study used a prespecified DAG-guided backdoor-adjustment framework to evaluate 16 DXA-derived hip skeletal phenotypes in relation to hip fracture risk in the UK Biobank. Under this framework, all evaluated phenotypes showed negative backdoor-adjusted ATEs per SD increase, with total femur BMC and total femur BMD showing the largest inverse ATEs. Heterogeneity analysis suggested that CATEs for total femur BMD varied across participant profiles, with more negative CATEs among older participants and those with lower BMI. In supplementary predictive evaluation, combining clinical variables with ATE-ranked skeletal phenotypes improved discrimination compared with FRAX in this analytic cohort. These findings suggest that phenotype-level causal evaluation of DXA-derived skeletal measures may help identify clinically informative features for hip fracture risk stratification.

## Statements and Declarations

### Funding

This research was supported in part by grants from the National Institutes of Health (U19AG055373, R01AG061917, P20GM109036, 1R15HL172198, and 1R15HL173852), the American Heart Association (#25AIREA1377168), and the Blue Cross Blue Shield of Michigan Foundation under grant number 2025110031.SAP.

### Acknowledgements

This research has been conducted using the UK Biobank Resource under Application Number 61915. The authors acknowledge the UK Biobank participants and staff for providing this research resource.

### Competing interests

The authors declare no competing interests.

### Data availability

The data used in this study cannot be shared publicly due to third-party data use restrictions and participant confidentiality requirements. Data are available to approved researchers through the UK Biobank application process:
https://www.ukbiobank.ac.uk/enable-your-research/about-our-data

### Code availability

The analysis code used in this study is available at:
https://github.com/MIILab-MTU/DXAAndHipFractureRiskByBackddoorCausalAnalysis

### Ethics approval

UK Biobank received ethical approval from the North West Multi-centre Research Ethics Committee.

### Consent to participate

All UK Biobank participants provided informed consent at recruitment.

**References:**

[1] Parker M, Johansen A. Hip fracture. Bmj. 2006;333(7557):27–30. doi: 10.1136/bmj.333.7557.27. PubMed PMID: 16809710; PubMed Central PMCID: PMC1488757.

[2] Shaik A, Larsen K, Lane NE, Zhao C, Su KJ, Keyak JH, Tian Q, Sha Q, Shen H, Deng HW, Zhou W. A staged approach using machine learning and uncertainty quantification to predict the risk of hip fracture. Bone Rep. 2024;22:101805. Epub 20240912. doi: 10.1016/j.bonr.2024.101805. PubMed PMID: 39328352; PubMed Central PMCID: PMC11426051.

[3] Sing CW, Lin TC, Bartholomew S, Bell JS, Bennett C, Beyene K, Bosco-Levy P, Bradbury BD, Chan AHY, Chandran M, Cooper C, de Ridder M, Doyon CY, Droz-Perroteau C, Ganesan G, Hartikainen S, Ilomaki J, Jeong HE, Kiel DP, Kubota K, Lai EC, Lange JL, Lewiecki EM, Lin J, Liu J, Maskell J, de Abreu MM, O'Kelly J, Ooba N, Pedersen AB, Prats-Uribe A, Prieto-Alhambra D, Qin SX, Shin JY, Sørensen HT, Tan KB, Thomas T, Tolppanen AM, Verhamme KMC, Wang GH, Watcharathanakij S, Wood SJ, Cheung CL, Wong ICK. Global Epidemiology of Hip Fractures: Secular Trends in Incidence Rate, Post-Fracture Treatment, and All-Cause Mortality. J Bone Miner Res. 2023;38(8):1064–75. Epub 20230529. doi: 10.1002/jbmr.4821. PubMed PMID: 37118993.

[4] Papadimitriou N, Tsilidis KK, Orfanos P, Benetou V, Ntzani EE, Soerjomataram I, Künn-Nelen A, Pettersson-Kymmer U, Eriksson S, Brenner H, Schöttker B, Saum KU, Holleczek B, Grodstein FD, Feskanich D, Orsini N, Wolk A, Bellavia A, Wilsgaard T, Jørgensen L, Boffetta P, Trichopoulos D, Trichopoulou A. Burden of hip fracture using disability-adjusted life-years: a pooled analysis of prospective cohorts in the CHANCES consortium. Lancet Public Health. 2017;2(5):e239–e46. Epub 20170411. doi: 10.1016/s2468-2667(17)30046-4. PubMed PMID: 29253489.
[5] Webster J, Oguzman E, Morris EJA, Shepperd S, Griffin XL, Johansen A, Goldacre R. Trends and variation in the incidence of hip fracture in England before, during, and after the COVID-19 pandemic (2014-2024): a population-based observational study. Lancet Reg Health Eur. 2025;57:101427. Epub 20250813. doi: 10.1016/j.lanepe.2025.101427. PubMed PMID: 40837274; PubMed Central PMCID: PMC12362259.
[6] Slart R, Punda M, Ali DS, Bazzocchi A, Bock O, Camacho P, Carey JJ, Colquhoun A, Compston J, Engelke K, Erba PA, Harvey NC, Krueger D, Lems WF, Lewiecki EM, Morgan S, Moseley KF, O'Brien C, Probyn L, Rhee Y, Richmond B, Schousboe JT, Shuhart C, Ward KA, Van den Wyngaert T, Zhang-Yin J, Khan AA. Updated practice guideline for dual-energy X-ray absorptiometry (DXA). Eur J Nucl Med Mol Imaging. 2025;52(2):539–63. Epub 20240924. doi: 10.1007/s00259-024-06912-6. PubMed PMID: 39316095; PubMed Central PMCID: PMC11732917.
[7] LeBoff MS, Greenspan SL, Insogna KL, Lewiecki EM, Saag KG, Singer AJ, Siris ES. The clinician's guide to prevention and treatment of osteoporosis. Osteoporos Int. 2022;33(10):2049–102. Epub 20220428. doi: 10.1007/s00198-021-05900-y. PubMed PMID: 35478046; PubMed Central PMCID: PMC9546973.
[8] Curry SJ, Krist AH, Owens DK, Barry MJ, Caughey AB, Davidson KW, Doubeni CA, Epling JW, Jr., Kemper AR, Kubik M, Landefeld CS, Mangione CM, Phipps MG, Pignone M, Silverstein M, Simon MA, Tseng CW, Wong JB. Screening for Osteoporosis to Prevent Fractures: US Preventive Services Task Force Recommendation Statement. Jama. 2018;319(24):2521–31. doi: 10.1001/jama.2018.7498. PubMed PMID: 29946735.
[9] Reid IR, McClung MR. Osteopenia: a key target for fracture prevention. Lancet Diabetes Endocrinol. 2024;12(11):856–64. Epub 20240923. doi: 10.1016/s2213-8587(24)00225-0. PubMed PMID: 39326428.
[10] Schini M, Johansson H, Harvey NC, Lorentzon M, Kanis JA, McCloskey EV. An overview of the use of the fracture risk assessment tool (FRAX) in osteoporosis. J Endocrinol Invest. 2024;47(3):501–11. Epub 20231024. doi: 10.1007/s40618-023-02219-9. PubMed PMID: 37874461; PubMed Central PMCID: PMC10904566.
[11] Bycroft C, Freeman C, Petkova D, Band G, Elliott LT, Sharp K, Motyer A, Vukcevic D, Delaneau O, O'Connell J, Cortes A, Welsh S, Young A, Effingham M, McVean G, Leslie S, Allen N, Donnelly P, Marchini J. The UK Biobank resource with deep phenotyping and genomic data. Nature. 2018;562(7726):203–9. Epub 20181010. doi: 10.1038/s41586-018-0579-z. PubMed PMID: 30305743; PubMed Central PMCID: PMC6786975.
[12] Wu Y, Chao J, Bao M, Zhang N. Predictive value of machine learning on fracture risk in osteoporosis: a systematic review and meta-analysis. BMJ Open. 2023;13(12):e071430. Epub 20231209. doi: 10.1136/bmjopen-2022-071430. PubMed PMID: 38070927; PubMed Central PMCID: PMC10728980.
[13] Qian Y, Xia J, Liu KQ, Xu L, Xie SY, Chen GB, Cong PK, Khederzadeh S, Zheng HF. Observational and genetic evidence highlight the association of human sleep behaviors with the incidence of fracture. Commun Biol. 2021;4(1):1339. Epub 20211126. doi: 10.1038/s42003-021-02861-0. PubMed PMID: 34837057; PubMed Central PMCID: PMC8626439.
[14] Kim Y, Kim YG, Park JW, Kim BW, Shin Y, Kong SH, Kim JH, Lee YK, Kim SW, Shin CS. A CT-based Deep Learning Model for Predicting Subsequent Fracture Risk in Patients with Hip Fracture. Radiology. 2024;310(1):e230614. doi: 10.1148/radiol.230614. PubMed PMID: 38289213.
[15] Yuan J, Li B, Zhang C, Wang J, Huang B, Ma L. Machine Learning-Based CT Radiomics Model to Predict the Risk of Hip Fragility Fracture. Acad Radiol. 2025;32(5):2854–62. Epub 20250203. doi: 10.1016/j.acra.2025.01.023. PubMed PMID: 39904664.
[16] Nethander M, Coward E, Reimann E, Grahnemo L, Gabrielsen ME, Wibom C, Mägi R, Funck-Brentano T, Hoff M, Langhammer A, Pettersson-Kymmer U, Hveem K, Ohlsson C. Assessment of the genetic and clinical determinants of hip fracture risk: Genome-wide association and Mendelian randomization study. Cell Rep Med. 2022;3(10):100776. doi: 10.1016/j.xcrm.2022.100776. PubMed PMID: 36260985; PubMed Central PMCID: PMC9589021.

[17] Ito C, Al-Hassany L, Kurth T, Glatz T. Distinguishing Description, Prediction, and Causal Inference: A Primer on Improving Congruence Between Research Questions and Methods. Neurology. 2025;104(4):e210171. Epub 20250203. doi: 10.1212/wnl.0000000000210171. PubMed PMID: 39899793.
[18] Dyer BP. The distinction between causal, predictive, and descriptive research-there is still room for improvement. J Clin Epidemiol. 2026;189:111960. Epub 20250901. doi: 10.1016/j.jclinepi.2025.111960. PubMed PMID: 40902863.
[19] Morin SN, Leslie WD, Schousboe JT. Osteoporosis: A Review. Jama. 2025;334(10):894–907. doi: 10.1001/jama.2025.6003. PubMed PMID: 40587168.
[20] Dahabreh IJ, Bibbins-Domingo K. Causal Inference About the Effects of Interventions From Observational Studies in Medical Journals. Jama. 2024;331(21):1845–53. doi: 10.1001/jama.2024.7741. PubMed PMID: 38722735.
[21] Correia LCL, Mascarenhas RF, De Menezes FSC, Oliveira JSJ, Vaccarino V, Ross JS, Wallach JD. Confounder Selection in Observational Studies in High-Impact Medical and Epidemiological Journals. JAMA Netw Open. 2025;8(7):e2524176. Epub 20250701. doi: 10.1001/jamanetworkopen.2025.24176. PubMed PMID: 40711790.
[22] Aga R, Søgaard AJ, Holvik K, Hagen TP, Idland G, Jeyananthan A, Melhuus KR, Rognerud M, Samuelsen SO, Meyer HE. Effectiveness of a fall and fracture prevention pathway on hip fracture risk, need of permanent care and mortality: a controlled before-and-after study. Age Ageing. 2025;54(6). doi: 10.1093/ageing/afaf161. PubMed PMID: 40552367; PubMed Central PMCID: PMC12206085.
[23] Mann HB, Whitney DR. On a Test of Whether One of Two Random Variables Is Stochastically Larger than the Other. The Annals of Mathematical Statistics. 1947;18(1):50–60. doi: 10.1214/aoms/1177730491.
[24] Pearson K. X. On the Criterion That a Given System of Deviations from the Probable in the Case of a Correlated System of Variables Is Such That It Can Be Reasonably Supposed to Have Arisen from Random Sampling. The London, Edinburgh, and Dublin Philosophical Magazine and Journal of Science. 1900;50(302):157–75. doi: 10.1080/14786440009463897.
[25] Fisher RA. On the Interpretation of chi-square from Contingency Tables, and the Calculation of P. Journal of the Royal Statistical Society. 1922;85(1):87–94. doi: 10.2307/2340521.
[26] Austin PC. Balance diagnostics for comparing the distribution of baseline covariates between treatment groups in propensity-score matched samples. Stat Med. 2009;28(25):3083–107. doi: 10.1002/sim.3697. PubMed PMID: 19757444; PubMed Central PMCID: PMC3472075.
[27] Feeney T, Hartwig FP, Davies NM. How to use directed acyclic graphs: guide for clinical researchers. Bmj. 2025;388:e078226. Epub 20250321. doi: 10.1136/bmj-2023-078226. PubMed PMID: 40118502.
[28] Jepsen KJ, Bredbenner TL, Karvonen-Gutierrez CA, Leis AM, Hood MM, Harlow SD, Randolph J, Clines GA, Merillat S, Elliott MR, Cauley JA, Greendale GA, Karlamangla AS, Peters KW, Harrison SL, Lui LY, Cawthon PM, Orwoll E. Femoral neck width is associated with unique trajectories of age-related hip structural changes and fracture risk within populations of adult women and men. J Bone Miner Res. 2025;40(10):1114–26. doi: 10.1093/jbmr/zjaf090. PubMed PMID: 40580057; PubMed Central PMCID: PMC12487784.
[29] Ensrud KE, Schousboe JT, Crandall CJ, Leslie WD, Fink HA, Cawthon PM, Kado DM, Lane NE, Cauley JA, Langsetmo L. Hip Fracture Risk Assessment Tools for Adults Aged 80 Years and Older. JAMA Netw Open. 2024;7(6):e2418612. Epub 20240603. doi: 10.1001/jamanetworkopen.2024.18612. PubMed PMID: 38941095; PubMed Central PMCID: PMC11214124.
[30] Marques EA, Carballido-Gamio J, Gudnason V, Sigurdsson G, Sigurdsson S, Aspelund T, Siggeirsdottir K, Launer L, Eiriksdottir G, Lang T, Harris TB. Sex differences in the spatial distribution of bone in relation to incident hip fracture: Findings from the AGES-Reykjavik study. Bone. 2018;114:72–80. Epub 20180516. doi: 10.1016/j.bone.2018.05.016. PubMed PMID: 29777918; PubMed Central PMCID: PMC6137723.
[31] Heymsfield SB, Hwaung P, Ferreyro-Bravo F, Heo M, Thomas DM, Schuna JM, Jr. Scaling of adult human bone and skeletal muscle mass to height in the US population. Am J Hum Biol. 2019;31(4):e23252. Epub 20190514. doi: 10.1002/ajhb.23252. PubMed PMID: 31087593; PubMed Central PMCID: PMC6634976.
[32] Xiao Z, Ren D, Feng W, Chen Y, Kan W, Xing D. Height and Risk of Hip Fracture: A Meta-Analysis of Prospective Cohort Studies. Biomed Res Int. 2016;2016:2480693. Epub 20161012. doi: 10.1155/2016/2480693. PubMed PMID: 27818998; PubMed Central PMCID: PMC5080474.

[33] Rikkonen T, Sund R, Sirola J, Honkanen R, Poole KES, Kröger H. Obesity is associated with early hip fracture risk in postmenopausal women: a 25-year follow-up. Osteoporos Int. 2021;32(4):769–77. Epub 20201023. doi: 10.1007/s00198-020-05665-w. PubMed PMID: 33095419; PubMed Central PMCID: PMC8026440.
[34] Amin H, Khan MA, Bukhari M. Deprivation indices and their association with fragility fractures and bone density: evidence from a large observational cohort. Rheumatology (Oxford). 2026;65(1). doi: 10.1093/rheumatology/keaf550. PubMed PMID: 41105243; PubMed Central PMCID: PMC12862389.
[35] Li H, Wallin M, Barregard L, Sallsten G, Lundh T, Ohlsson C, Mellström D, Andersson EM. Smoking-Induced Risk of Osteoporosis Is Partly Mediated by Cadmium From Tobacco Smoke: The MrOS Sweden Study. J Bone Miner Res. 2020;35(8):1424–9. Epub 20200406. doi: 10.1002/jbmr.4014. PubMed PMID: 32191351.
[36] Zoulakis M, Ambjörn M, Jaiswal R, Axelsson KF, Litsne H, Johansson L, Lorentzon M. Impact of Current and Previous Smoking on Fracture Risk in Older Women: The Role of Physical Function, Bone Density and Bone Microarchitecture. J Bone Miner Res. 2026. Epub 20260210. doi: 10.1093/jbmr/zjag028. PubMed PMID: 41665281.
[37] Godos J, Giampieri F, Chisari E, Micek A, Paladino N, Forbes-Hernández TY, Quiles JL, Battino M, La Vignera S, Musumeci G, Grosso G. Alcohol Consumption, Bone Mineral Density, and Risk of Osteoporotic Fractures: A Dose-Response Meta-Analysis. Int J Environ Res Public Health. 2022;19(3). Epub 20220128. doi: 10.3390/ijerph19031515. PubMed PMID: 35162537; PubMed Central PMCID: PMC8835521.
[38] Denova-Gutiérrez E, Méndez-Sánchez L, Muñoz-Aguirre P, Tucker KL, Clark P. Dietary Patterns, Bone Mineral Density, and Risk of Fractures: A Systematic Review and Meta-Analysis. Nutrients. 2018;10(12). Epub 20181205. doi: 10.3390/nu10121922. PubMed PMID: 30563066; PubMed Central PMCID: PMC6316557.
[39] Ketabforoush A, Aleahmad M, Qorbani M, Mehrpoor G, Afrashteh S, Mardi S, Dolatshahi E. Bone mineral density status in patients with recent-onset rheumatoid arthritis. J Diabetes Metab Disord. 2023;22(1):775–85. Epub 20230317. doi: 10.1007/s40200-023-01200-w. PubMed PMID: 37250372; PubMed Central PMCID: PMC10023217.
[40] Kanis JA, Johansson H, McCloskey EV, Liu E, Schini M, Vandenput L, Åkesson KE, Anderson FA, Azagra R, Bager CL, Beaudart C, Bischoff-Ferrari HA, Biver E, Bruyère O, Cauley JA, Center JR, Chapurlat R, Christiansen C, Cooper C, Crandall CJ, Cummings SR, da Silva JAP, Dawson-Hughes B, Diez-Perez A, Dufour AB, Eisman JA, Elders PJM, Ferrari S, Fujita Y, Fujiwara S, Glüer CC, Goldshtein I, Goltzman D, Gudnason V, Hall J, Hans D, Hoff M, Hollick RJ, Huisman M, Iki M, Ish-Shalom S, Jones G, Karlsson MK, Khosla S, Kiel DP, Koh WP, Koromani F, Kotowicz MA, Kröger H, Kwok T, Lamy O, Langhammer A, Larijani B, Lippuner K, McGuigan FEA, Mellström D, Merlijn T, Nguyen TV, Nordström A, Nordström P, TW ON, Obermayer-Pietsch B, Ohlsson C, Orwoll ES, Pasco JA, Rivadeneira F, Schott AM, Shiroma EJ, Siggeirsdottir K, Simonsick EM, Sornay-Rendu E, Sund R, Swart K, Szulc P, Tamaki J, Torgerson DJ, van Schoor NM, van Staa TP, Vila J, Wright NC, Yoshimura N, Zillikens MC, Zwart M, Harvey NC, Lorentzon M, Leslie WD. Rheumatoid arthritis and subsequent fracture risk: an individual person meta-analysis to update FRAX. Osteoporos Int. 2025;36(4):653–71. Epub 20250216. doi: 10.1007/s00198-025-07397-1. PubMed PMID: 39955689.
[41] Leslie WD, Lix LM, Binkley N. Osteoporosis treatment considerations based upon fracture history, fracture risk assessment, vertebral fracture assessment, and bone density in Canada. Arch Osteoporos. 2020;15(1):93. Epub 20200623. doi: 10.1007/s11657-020-00775-8. PubMed PMID: 32577922.
[42] Hu Q, Liu H, Du Y, Duan R, Li L, Yang D, Ouyang Z. Walking pace and its association with osteoporosis and pathological fractures: insights from UK biobank. Front Endocrinol (Lausanne). 2025;16:1635999. Epub 20250826. doi: 10.3389/fendo.2025.1635999. PubMed PMID: 40933376; PubMed Central PMCID: PMC12417151.
[43] Huang SG, Lee RP, Yao TK, Wang JH, Wu WT, Yeh KT. Correlation Between Handgrip Strength and Bone Density and Fragility Fracture Risk Among Older Adults: A Cross-Sectional Study. J Nurs Res. 2025;33(1):e375. Epub 20250201. doi: 10.1097/jnr.0000000000000656. PubMed PMID: 39835766.
[44] Hernán MA, Robins JM. Causal Inference: What If. Boca Raton: Chapman & Hall/CRC; 2020.
[45] Pearl J. Causality: Models, Reasoning, and Inference. 2nd ed: Cambridge University Press; 2009.

[46] Athey S, Tibshirani J, Wager S. Generalized Random Forests. The Annals of Statistics. 2019;47(2):1148–78. doi: 10.1214/18-AOS1709.
[47] Wager S, Athey S. Estimation and Inference of Heterogeneous Treatment Effects using Random Forests. Journal of the American Statistical Association. 2018;113(523):1228–42. doi: 10.1080/01621459.2017.1319839.
[48] Hsieh CI, Zheng K, Lin C, Mei L, Lu L, Li W, Chen FP, Wang Y, Zhou X, Wang F, Xie G, Xiao J, Miao S, Kuo CF. Automated bone mineral density prediction and fracture risk assessment using plain radiographs via deep learning. Nat Commun. 2021;12(1):5472. Epub 20210916. doi: 10.1038/s41467-021-25779-x. PubMed PMID: 34531406; PubMed Central PMCID: PMC8446034.
[49] Jaiswal R, Pivodic A, Zoulakis M, Axelsson KF, Litsne H, Johansson L, Lorentzon M. Prediction of hip fracture by high-resolution peripheral quantitative computed tomography in older Swedish women. J Bone Miner Res. 2025;40(6):779–90. doi: 10.1093/jbmr/zjaf020. PubMed PMID: 39869791; PubMed Central PMCID: PMC12131241.
[50] Tobias JH, Nethander M, Faber BG, Heppenstall SV, Ebsim R, Cootes T, Lindner C, Saunders FR, Gregory JS, Aspden RM, Harvey NC, Kemp JP, Frysz M, Ohlsson C. Femoral neck width genetic risk score is a novel independent risk factor for hip fractures. J Bone Miner Res. 2024;39(3):241–51. doi: 10.1093/jbmr/zjae002. PubMed PMID: 38477772; PubMed Central PMCID: PMC11240160.
[51] Künzel SR, Sekhon JS, Bickel PJ, Yu B. Metalearners for estimating heterogeneous treatment effects using machine learning. Proc Natl Acad Sci U S A. 2019;116(10):4156–65. Epub 20190215. doi: 10.1073/pnas.1804597116. PubMed PMID: 30770453; PubMed Central PMCID: PMC6410831.
[52] Khan AA, Slart R, Ali DS, Bock O, Carey JJ, Camacho P, Engelke K, Erba PA, Harvey NC, Lems WF, Morgan S, Moseley KF, O'Brien C, Probyn L, Punda M, Richmond B, Schousboe JT, Shuhart C, Ward KA, Lewiecki EM. Osteoporotic Fractures: Diagnosis, Evaluation, and Significance From the International Working Group on DXA Best Practices. Mayo Clin Proc. 2024;99(7):1127–41. doi: 10.1016/j.mayocp.2024.01.011. PubMed PMID: 38960497.
[53] Lundberg SM, Lee S-I, editors. A Unified Approach to Interpreting Model Predictions. Advances in Neural Information Processing Systems; 2017: Curran Associates, Inc.
[54] Li GH, Cheung CL, Tan KC, Kung AW, Kwok TC, Lau WC, Wong JS, Hsu WWQ, Fang C, Wong IC. Development and validation of sex-specific hip fracture prediction models using electronic health records: a retrospective, population-based cohort study. EClinicalMedicine. 2023;58:101876. Epub 20230227. doi: 10.1016/j.eclinm.2023.101876. PubMed PMID: 36896245; PubMed Central PMCID: PMC9989633.

# Supplementary

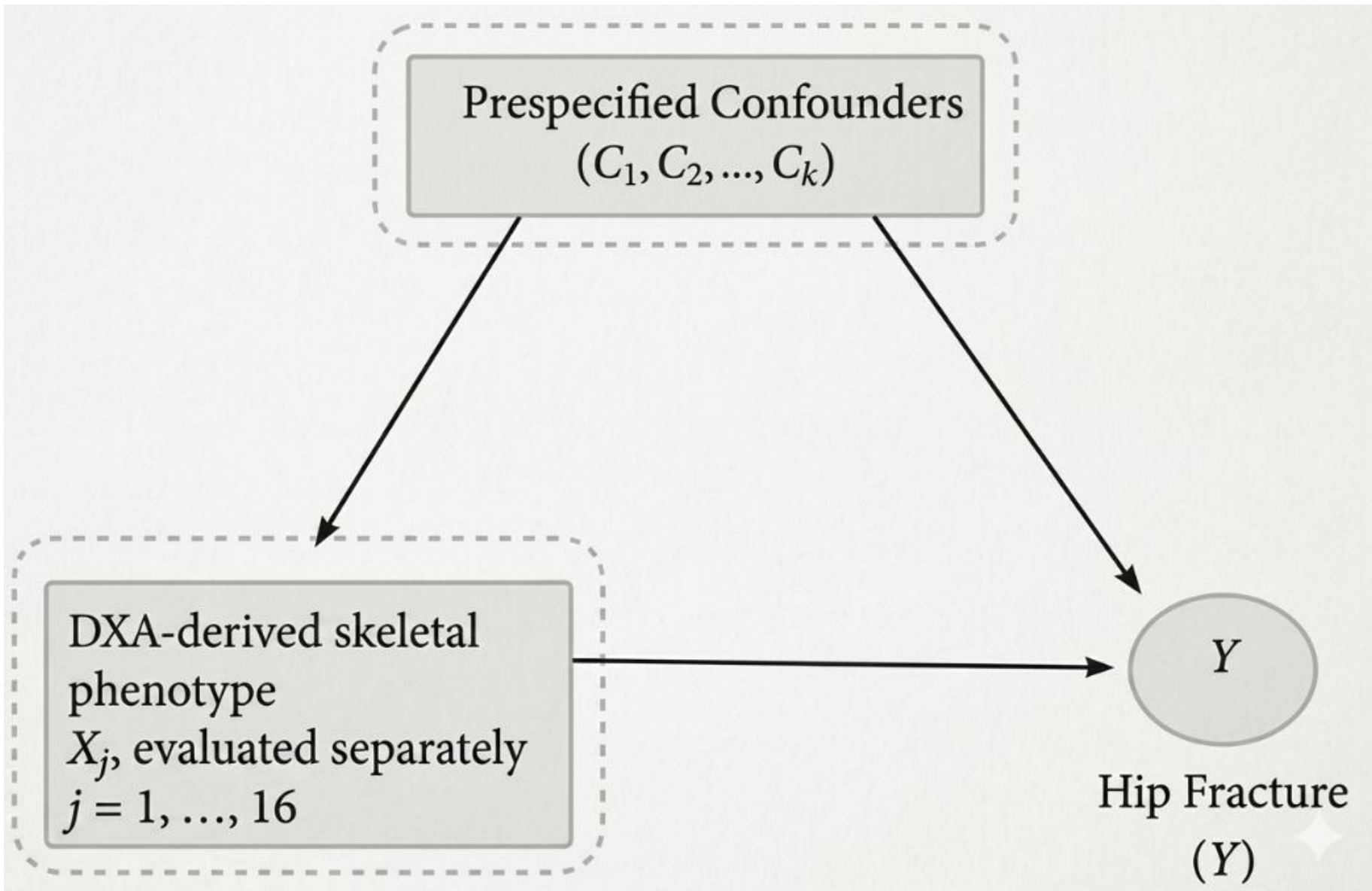

Supplementary Figure S1. Conceptual directed acyclic graph used to guide confounder selection for phenotype-specific backdoor-adjusted analyses. Each DXA-derived skeletal phenotype $X_j (j = 1, \ldots, 16)$was evaluated separately as the exposure, hip fracture was treated as the outcome $Y$, and the prespecified confounder set $C$was used to block major backdoor paths of the form $X_j \leftarrow C \rightarrow Y$.

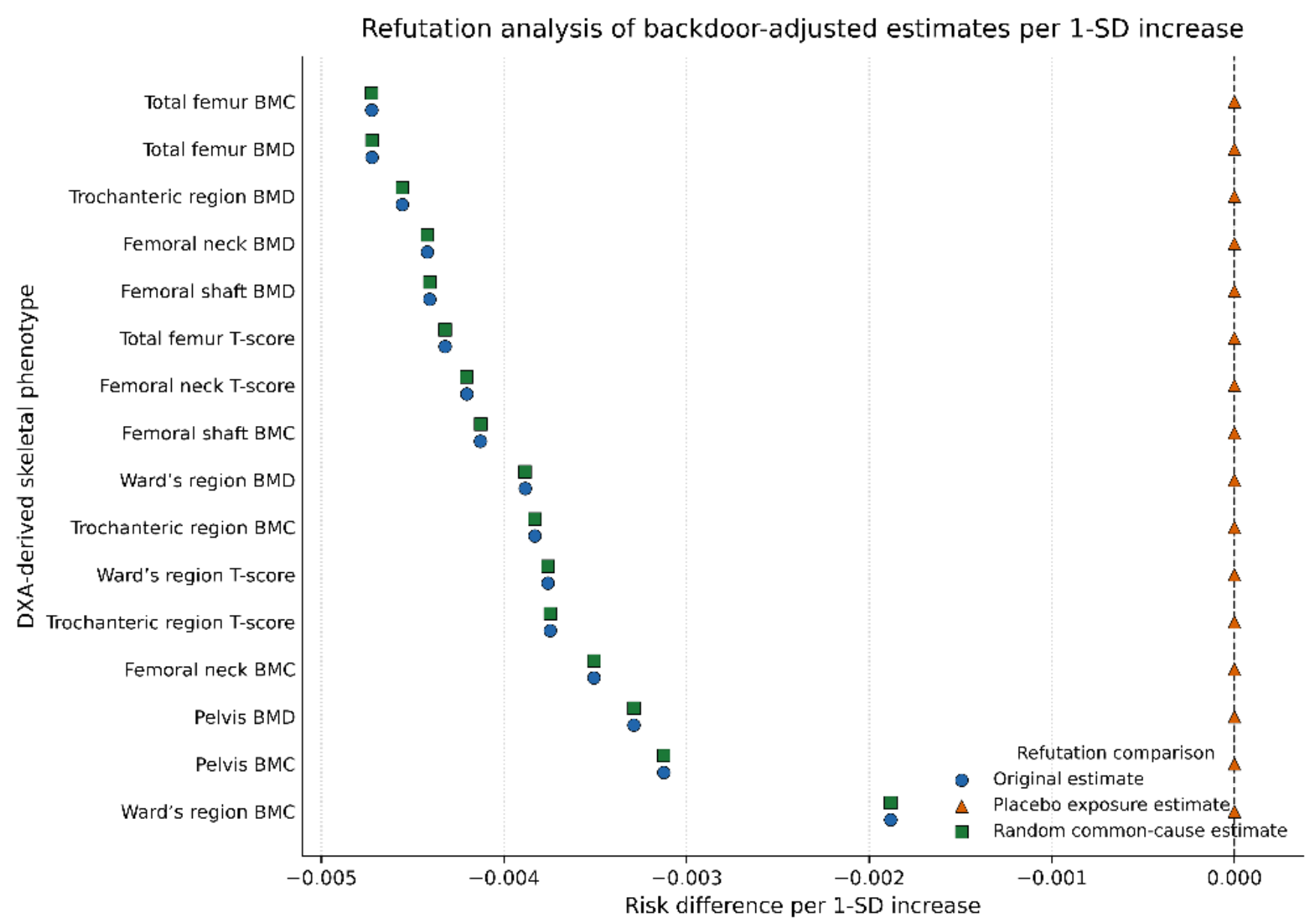

Supplementary Figure S2. Refutation analysis of backdoor-adjusted effect estimates per 1-SD increase.

All Treatments - CATE Feature Importance

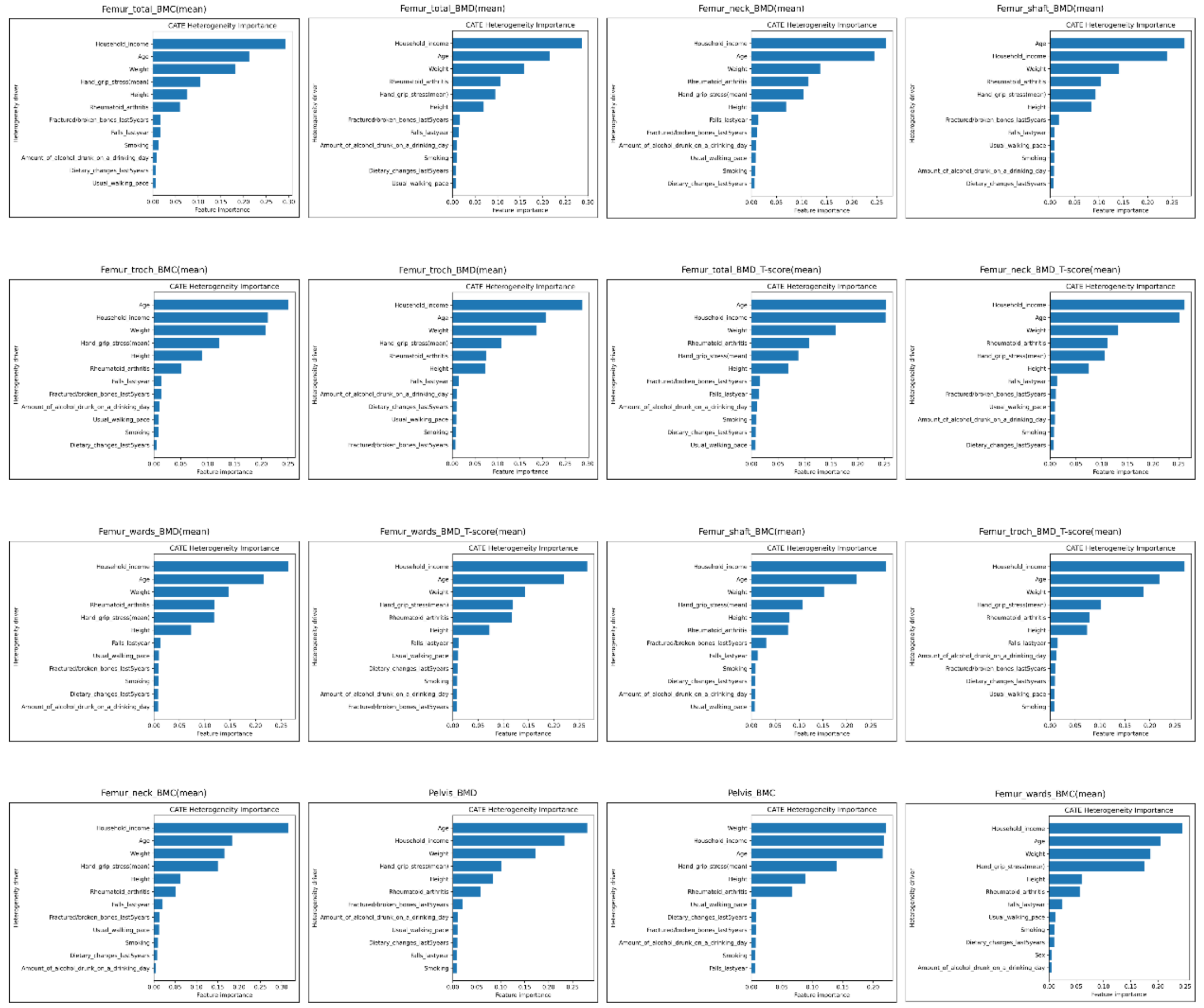

Supplementary Figure S3. Variable importance for CATE heterogeneity across all 16 DXA-derived skeletal phenotypes.

All Treatments - CATE Curve by Age and Sex

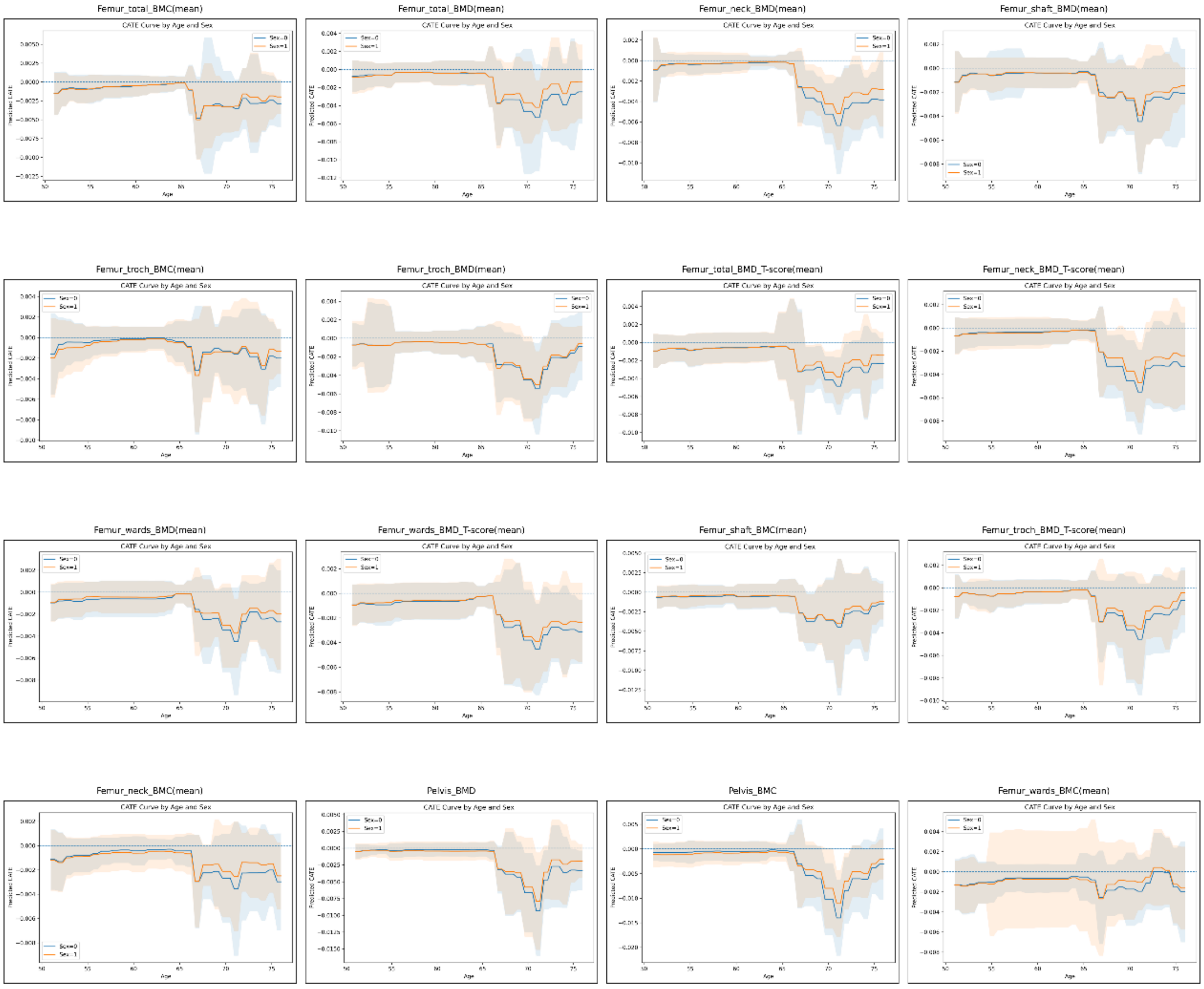

Supplementary Figure S4. Predicted CATE curves by age and sex across all 16 DXA-derived skeletal phenotypes.

All Treatments - CATE Decile Summary

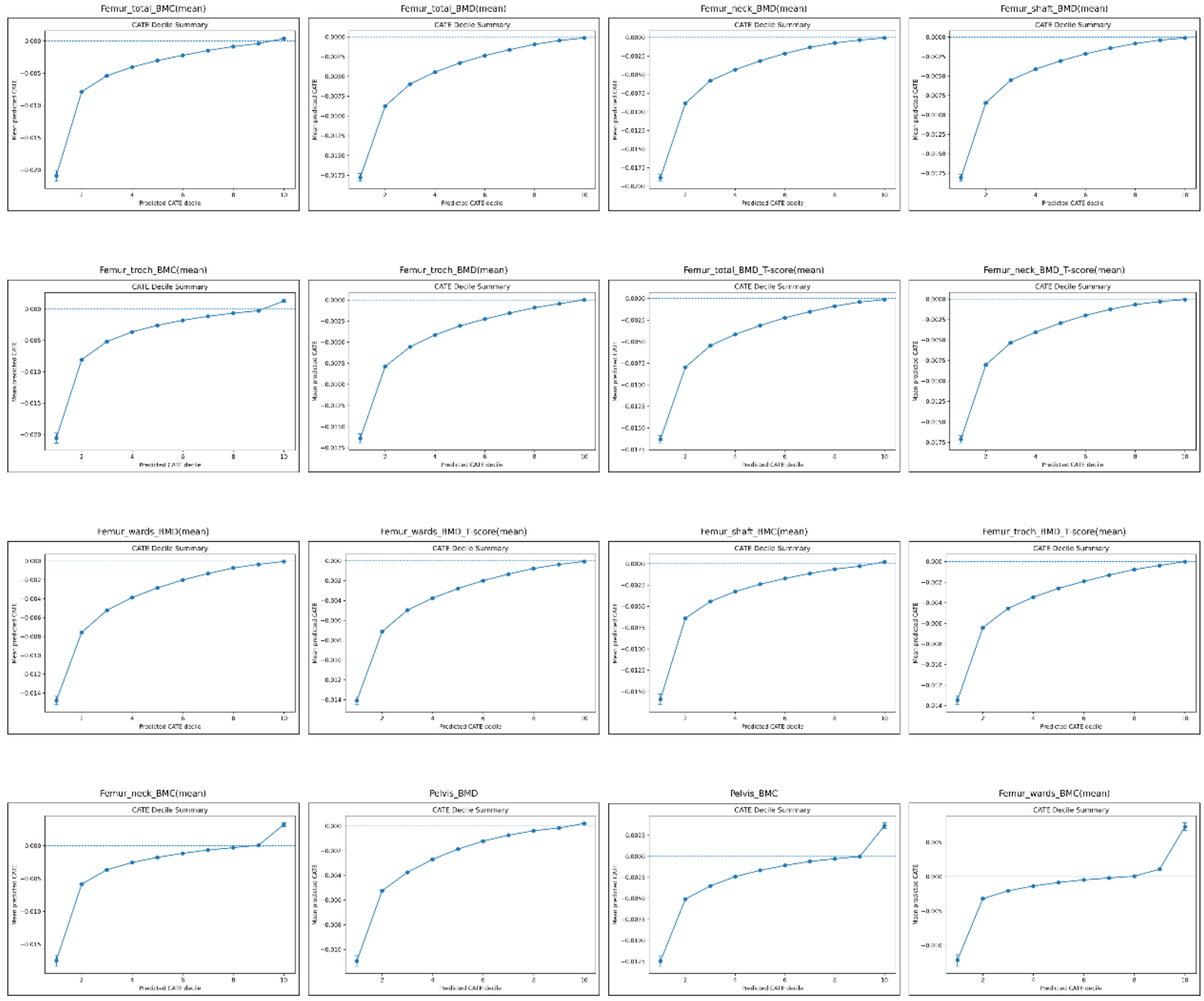

Supplementary Figure S5. Decile-based summary of mean predicted CATE across all 16 DXA-derived skeletal phenotypes.

All Treatments - Subgroup CATE by Age

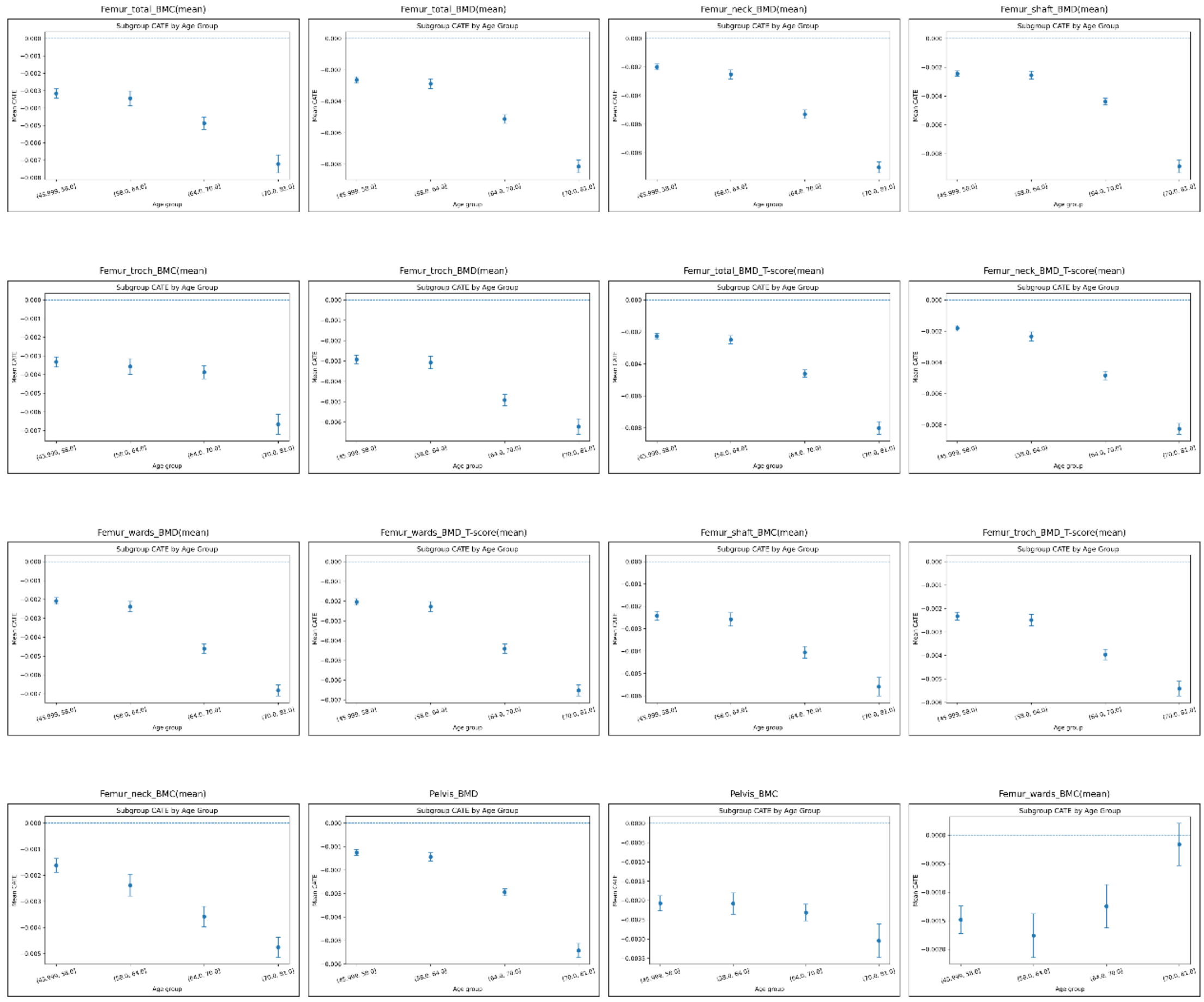

Supplementary Figure S6. Mean CATE by age group across all 16 DXA-derived skeletal phenotypes.

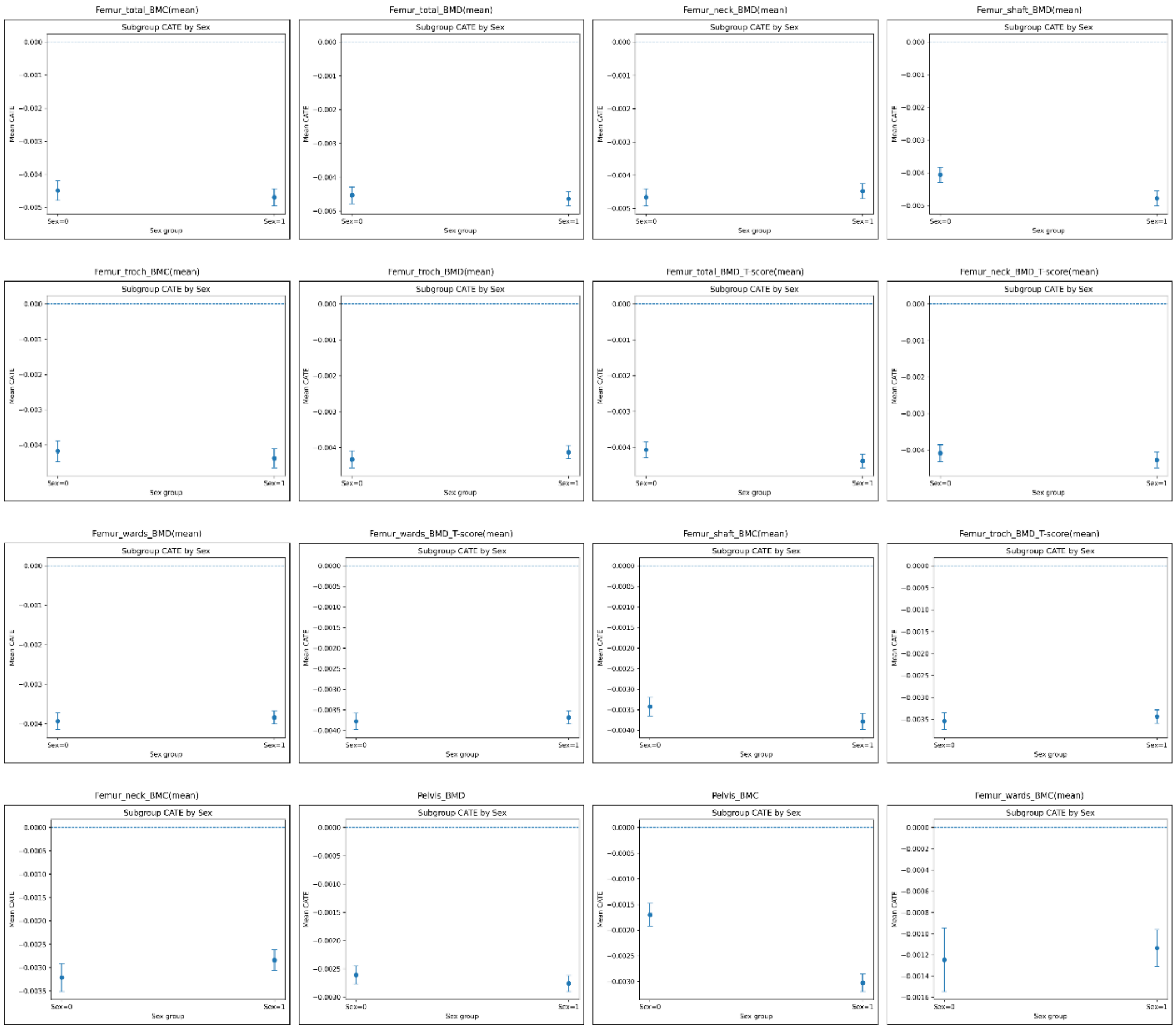

Supplementary Figure S7. Mean CATE by sex group across all 16 DXA-derived skeletal phenotypes.

All Treatments - Subgroup CATE by BMI

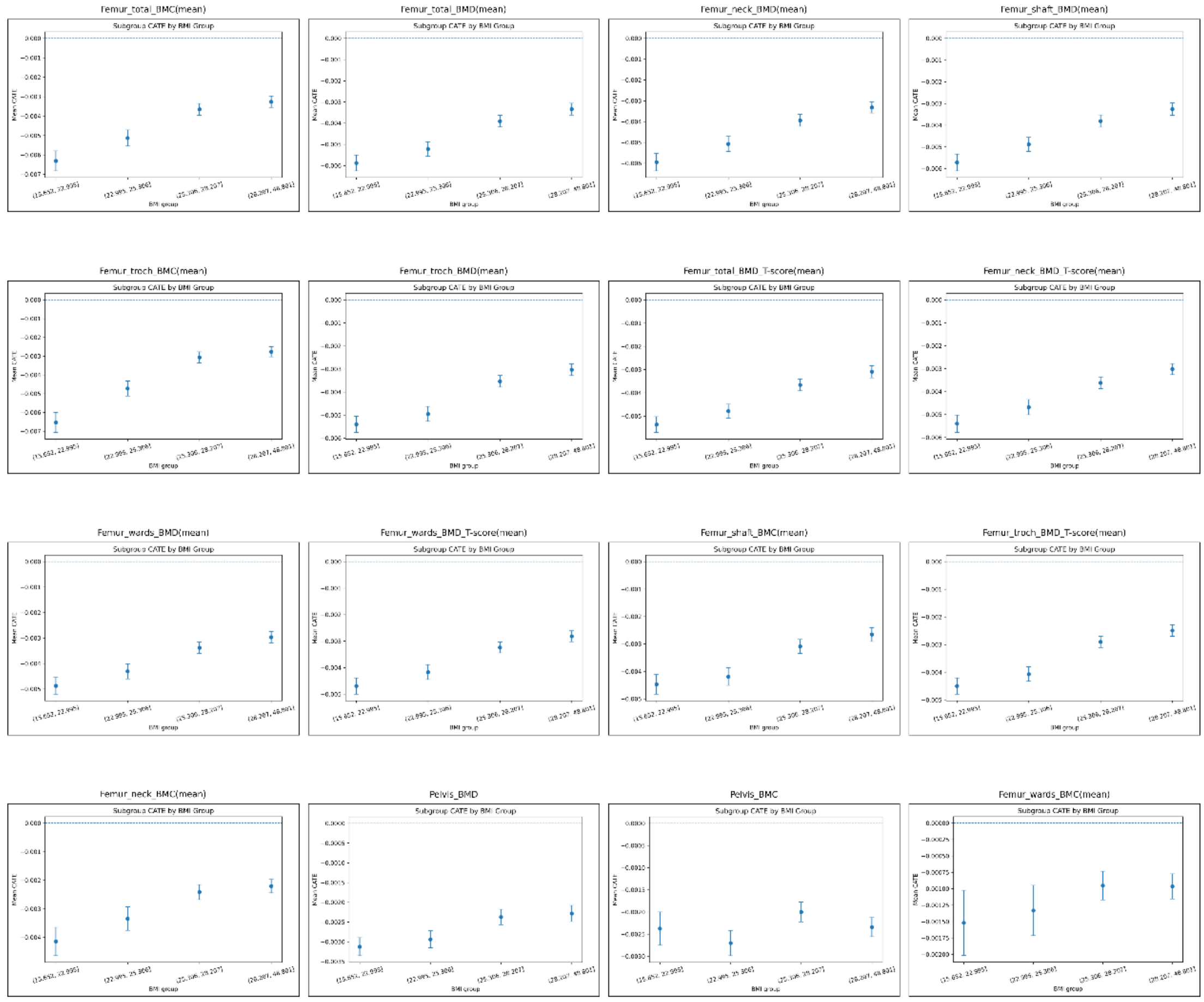

Supplementary Figure S8. Mean CATE by BMI group across all 16 DXA-derived skeletal phenotypes.

All Treatments - ITE Distribution

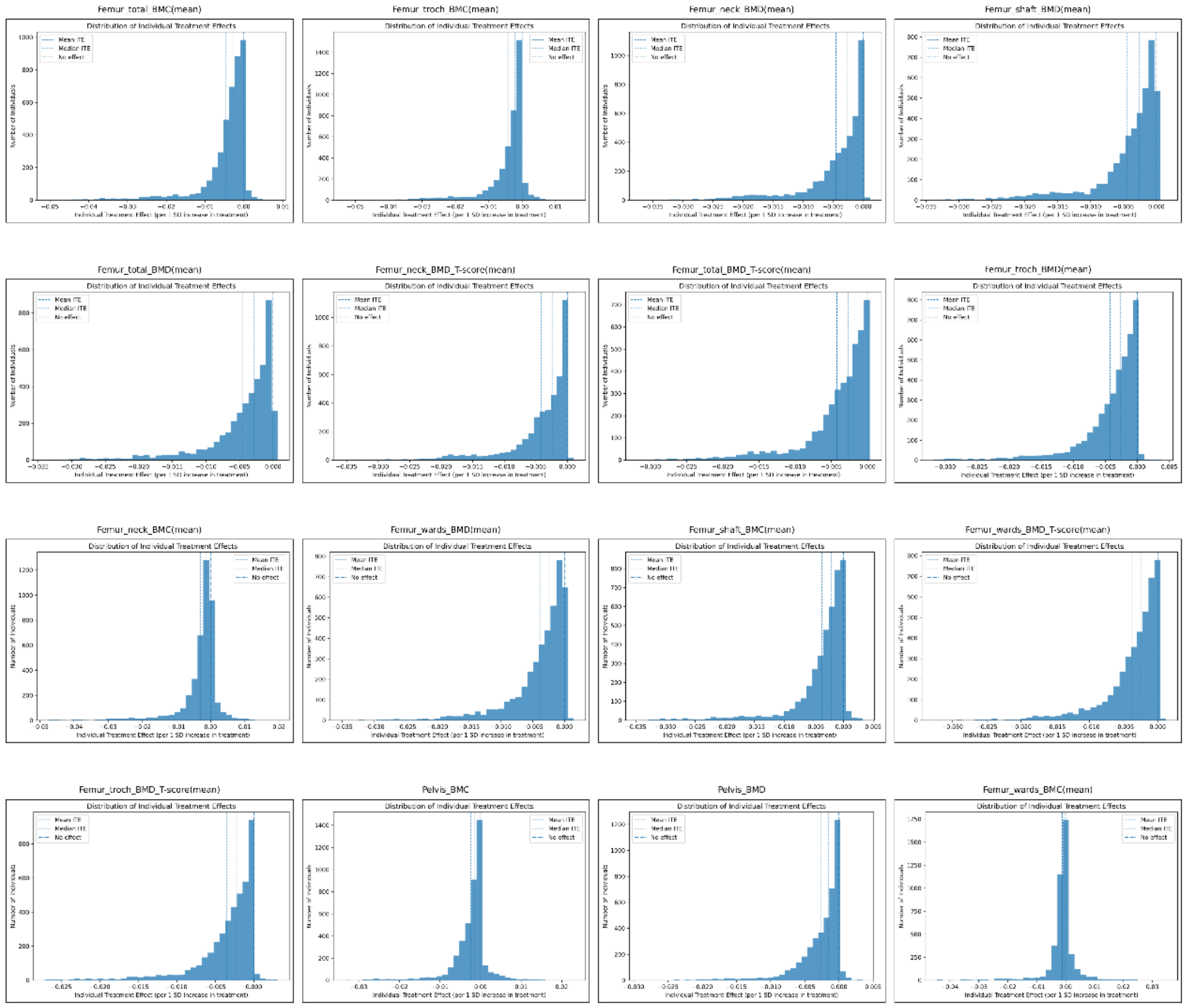

Supplementary Figure S9. Distribution of individual predicted treatment effects across all 16 DXA-derived skeletal phenotypes.

All Treatments - ITE Ranking Curve

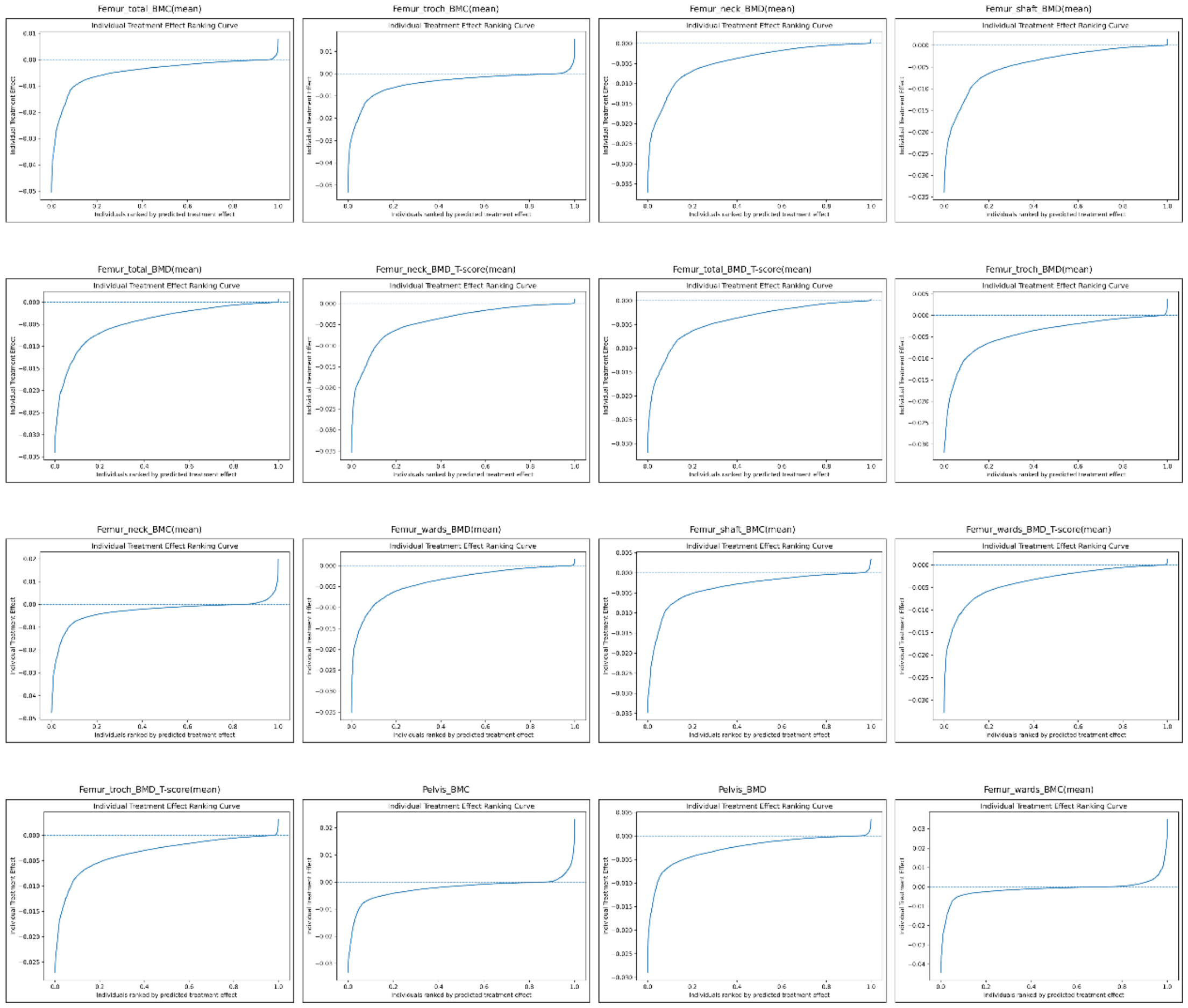

Supplementary Figure S10. Ranking curves of individual predicted treatment effects across all 16 DXA-derived skeletal phenotypes.

All Treatments - Top-Benefit Population

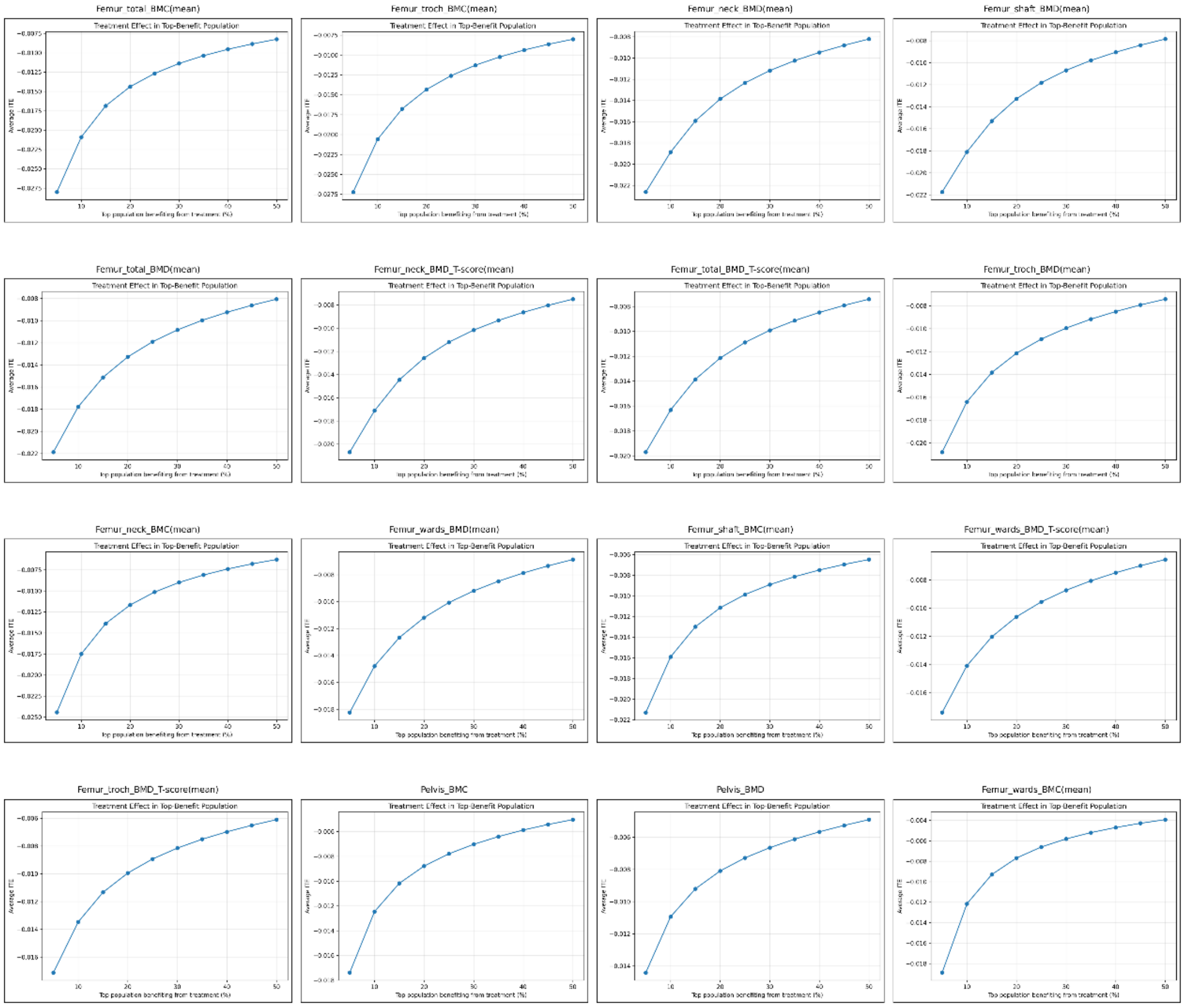

Supplementary Figure S11. Average predicted treatment effect in the top-benefit population across all 16 DXA-derived skeletal phenotypes.

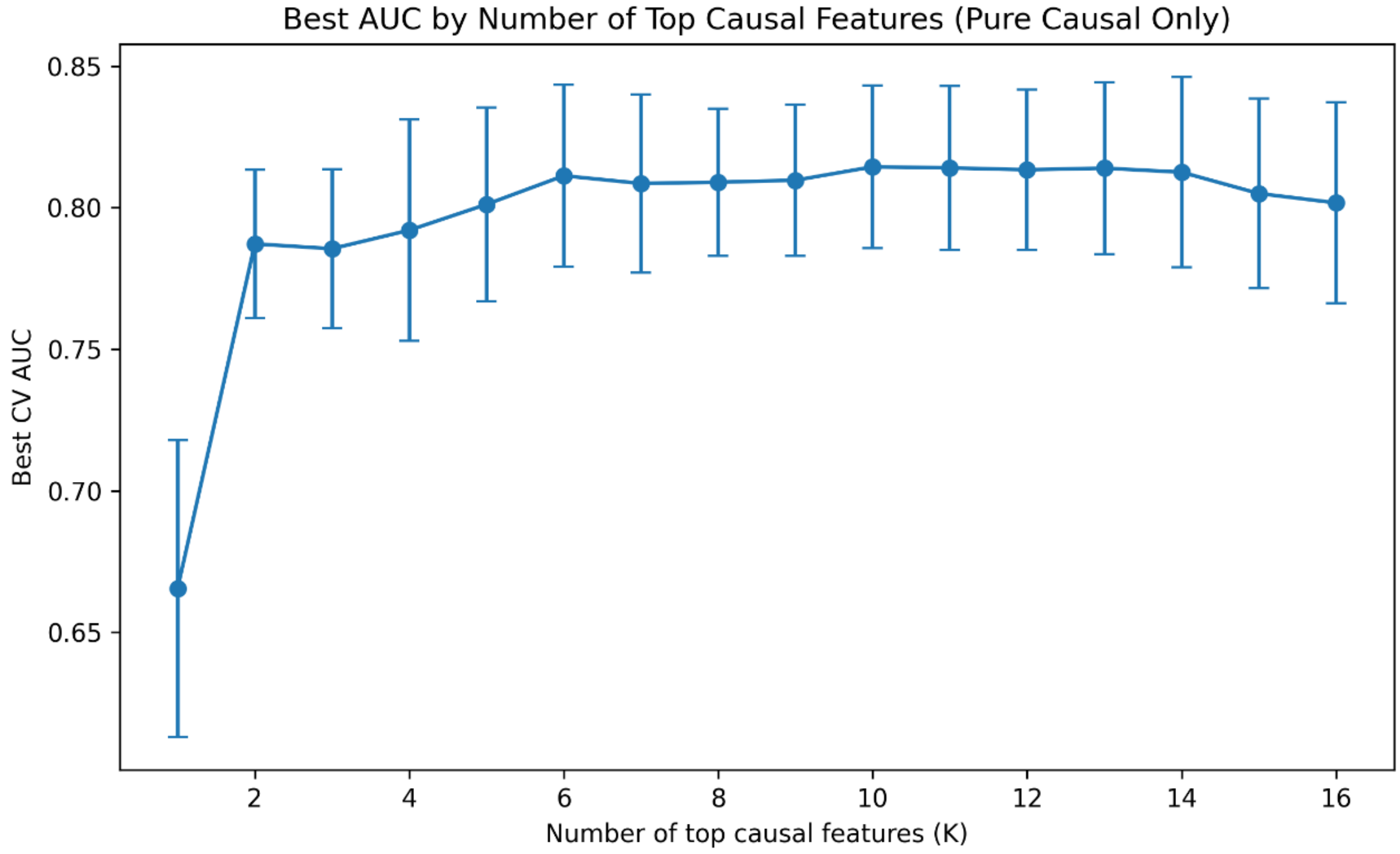

Supplementary Figure S12. Best cross-validated AUC by number of top-ranked causal skeletal phenotypes in the pure causal top-K analysis.

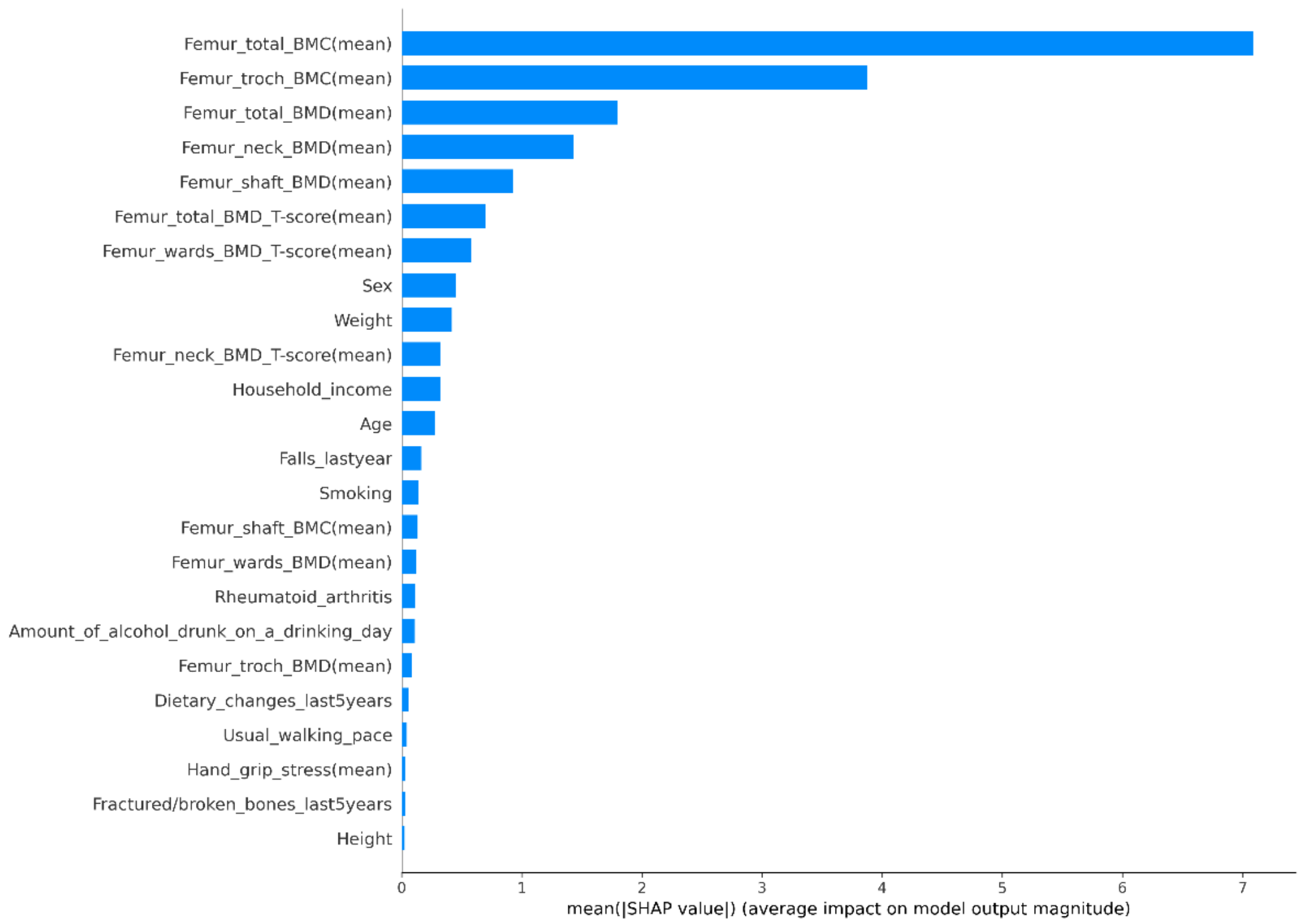

Supplementary Figure S13A. Global SHAP summary plot of feature importance in the final logistic regression model.

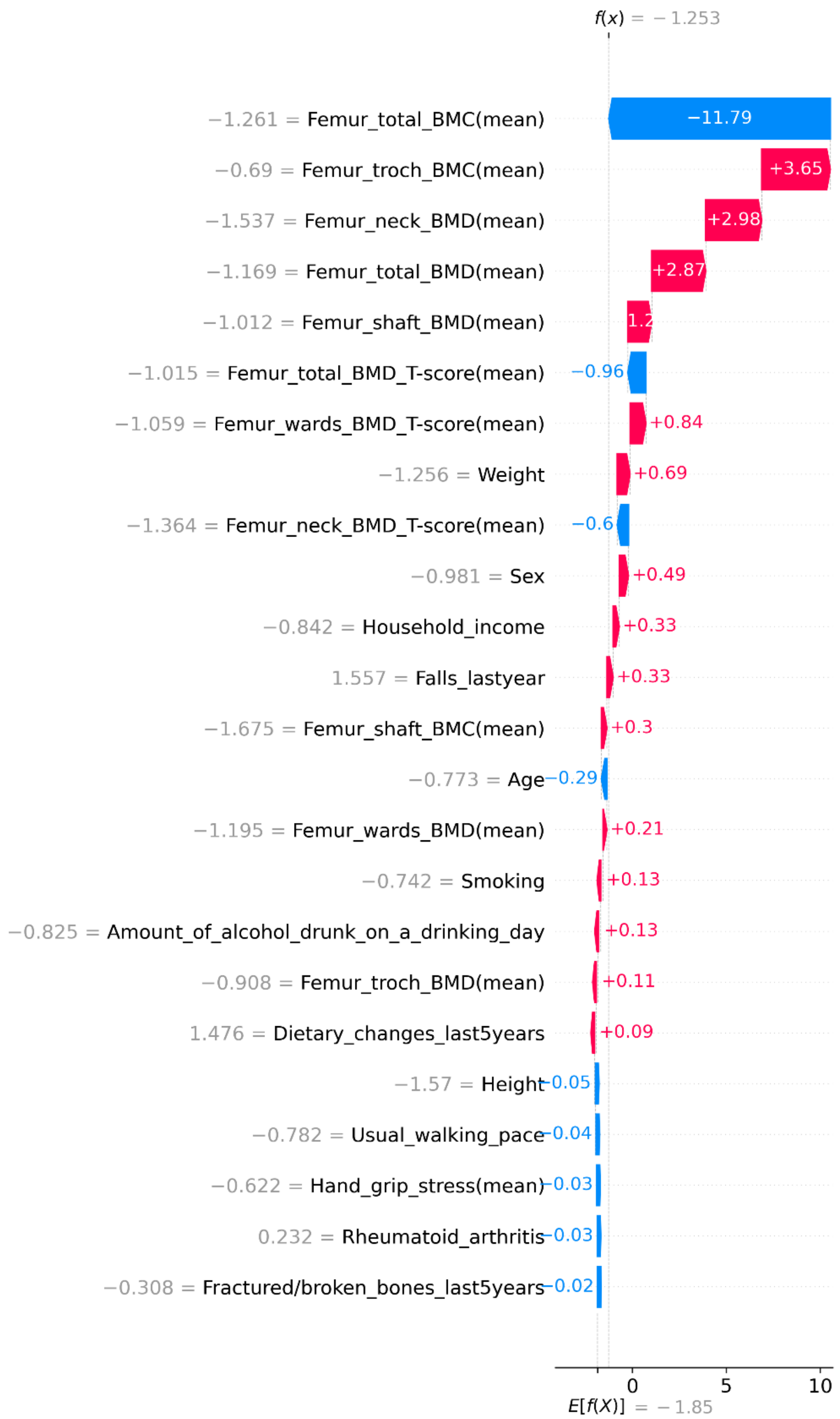

Supplementary Figure S13B. SHAP waterfall plot showing feature contributions for a representative individual prediction in the final logistic regression model.

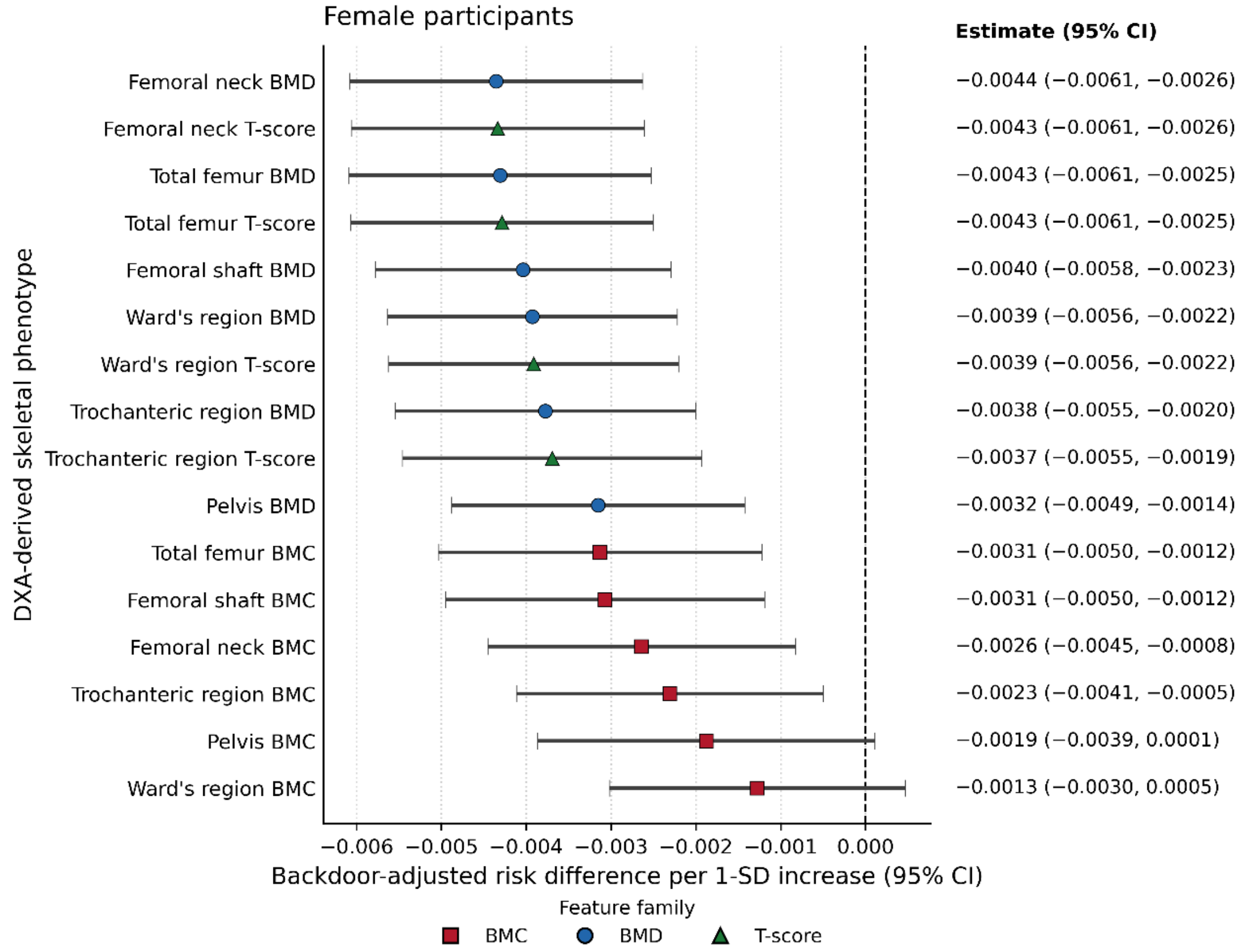

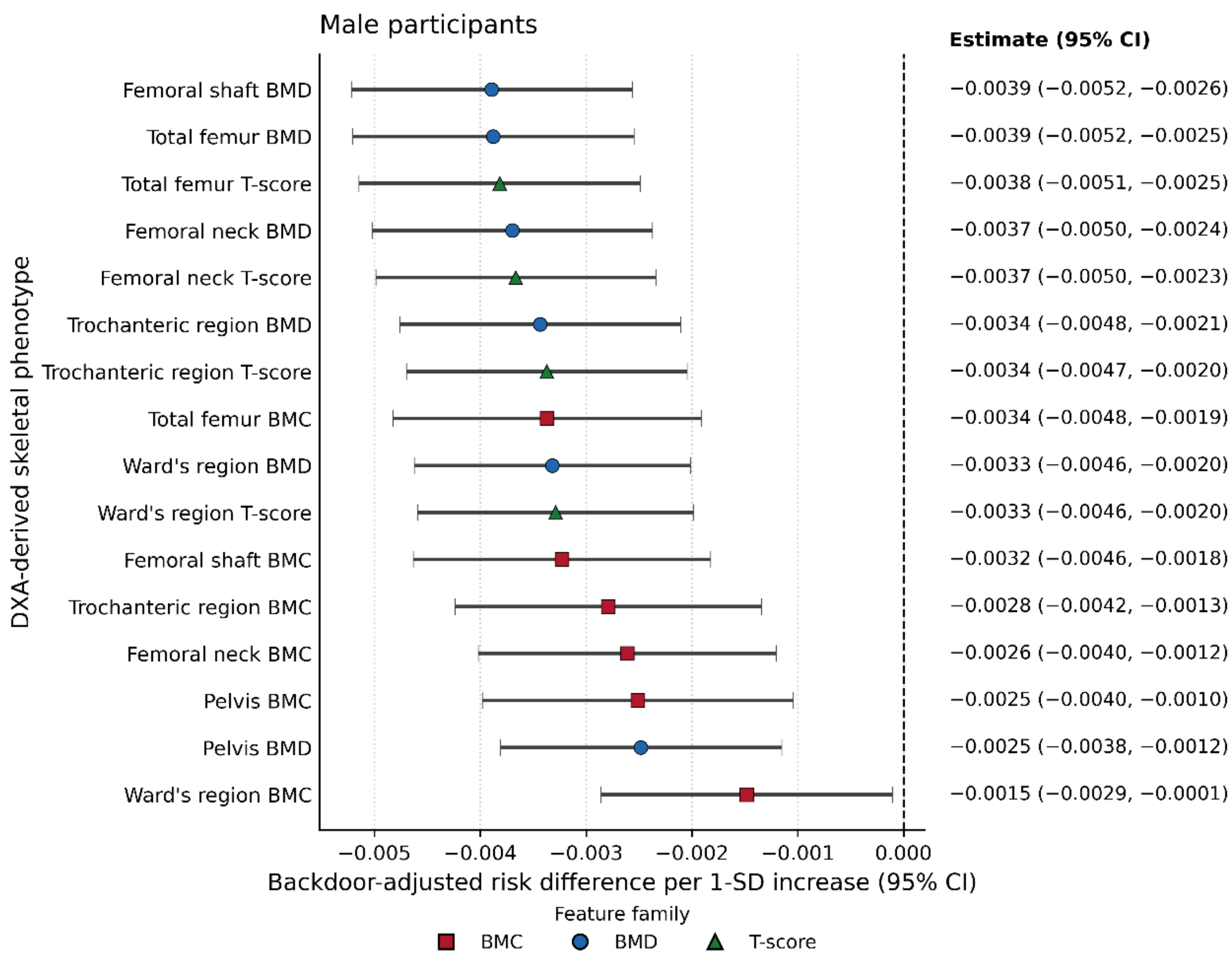

Supplementary Figure S14. Backdoor-adjusted causal effects of DXA-derived bone features on hip fracture risk stratified by sex.

Supplementary Table S1. Full backdoor-adjusted estimates for all 16 phenotypes

| X_variable | Family | Region | Estimate_perSD | CI95_low | CI95_high |
|---|---|---|---|---|---|
| Femur_total_BMC | BMC | total | -0.0047 | -0.0064 | -0.0031 |
| Femur_total_BMD | BMD | total | -0.0047 | -0.0059 | -0.0035 |
| Femur_troch_BMD | BMD | troch | -0.0046 | -0.0059 | -0.0032 |
| Femur_neck_BMD | BMD | neck | -0.0044 | -0.0056 | -0.0033 |
| Femur_shaft_BMD | BMD | shaft | -0.0044 | -0.0056 | -0.0033 |
| Femur_total_T-score | T-score | total | -0.0043 | -0.0054 | -0.0032 |
| Femur_neck_T-score | T-score | neck | -0.0042 | -0.0053 | -0.0031 |
| Femur_shaft_BMC | BMC | shaft | -0.0041 | -0.0056 | -0.0027 |
| Femur_wards_BMD | BMD | wards | -0.0039 | -0.0050 | -0.0028 |
| Femur_troch_BMC | BMC | troch | -0.0039 | -0.0055 | -0.0022 |
| Femur_wards_T-score | T-score | wards | -0.0038 | -0.0048 | -0.0027 |
| Femur_troch_T-score | T-score | troch | -0.0037 | -0.0048 | -0.0026 |
| Femur_neck_BMC | BMC | neck | -0.0035 | -0.0049 | -0.0021 |
| Pelvis_BMD | BMD | pelvis | -0.0033 | -0.0045 | -0.0021 |
| Pelvis_BMC | BMC | pelvis | -0.0031 | -0.0047 | -0.0016 |
| Femur_wards_BMC | BMC | wards | -0.0019 | -0.0033 | -0.0005 |

Supplementary Table S2. Performance comparison of the 10 evaluated machine-learning classifiers.

| Classifier | AUC | Accuracy | Sensitivity | Specificity | Selection status |
|---|---|---|---|---|---|
| LogisticRegression | 0.842 | 0.793 | 0.748 | 0.793 | Selected |
| GradientBoosting | 0.838 | 0.994 | 0.270 | 0.998 | Not selected |
| AdaBoost | 0.829 | 0.996 | 0.235 | 1 | Not selected |
| GaussianNB | 0.811 | 0.823 | 0.652 | 0.823 | Not selected |
| ExtraTrees | 0.787 | 0.996 | 0.226 | 1 | Not selected |
| RandomForest | 0.761 | 0.995 | 0.026 | 1 | Not selected |
| SVM_RBF | 0.737 | 0.995 | 0 | 1 | Not selected |
| MLP | 0.728 | 0.994 | 0.130 | 0.999 | Not selected |
| KNN | 0.656 | 0.995 | 0.183 | 1 | Not selected |
| DecisionTree | 0.571 | 0.991 | 0.148 | 0.995 | Not selected |

In the present comparison, class-weighted learning (class_weight="balanced") was implemented for LogisticRegression, RandomForest, ExtraTrees, SVM with RBF kernel, and DecisionTree, because these classifiers directly support class-weight specification in scikit-learn. By contrast, GradientBoosting, AdaBoost, KNN, GaussianNB, and MLP do not directly provide a class_weight constructor argument in the same way, so they were evaluated without this option. Several models nonetheless showed near-perfect specificity but low sensitivity, which likely reflects the marked class imbalance in the analytic cohort and a tendency to favor the majority class under the current evaluation setting. Logistic regression was selected as the final classifier because it achieved the best overall discrimination while also providing a more balanced and interpretable framework for downstream analyses.

Supplementary Table S3. Sex-stratified and full-cohort predictive performance of logistic-regression-based feature-set configurations.

| Sex | Dataset | AUC | ACC | Sen | Spe |
|---|---|---|---|---|---|
| Female | Clinical_only | 0.821 | 0.765 | 0.731 | 0.765 |
| | Top11_causal_only | 0.800 | 0.716 | 0.774 | 0.715 |
| | All_causal | 0.795 | 0.722 | 0.718 | 0.722 |
| | All_features | 0.860 | 0.820 | 0.748 | 0.821 |
| | Clinical_top11_causal | 0.863 | 0.815 | 0.748 | 0.816 |
| Male | Clinical_only | 0.676 | 0.703 | 0.528 | 0.704 |
| | Top8_causal_only | 0.752 | 0.722 | 0.658 | 0.722 |
| | All_causal | 0.702 | 0.735 | 0.594 | 0.735 |
| | All_features | 0.728 | 0.789 | 0.528 | 0.790 |
| | Clinical_top8_causal | 0.770 | 0.778 | 0.639 | 0.778 |
| All cohort | Clinical_only | 0.754 | 0.733 | 0.626 | 0.734 |
| | Top11_causal_only | 0.813 | 0.721 | 0.783 | 0.721 |
| | All_causal | 0.801 | 0.728 | 0.757 | 0.727 |
| | All_features | 0.837 | 0.797 | 0.713 | 0.797 |
| | Clinical_top11_causal | 0.842 | 0.793 | 0.748 | 0.793 |